\documentclass[11pt]{article}
\usepackage[margin=1in]{geometry}
\usepackage{amsmath,amssymb,amsfonts,bbm,mathrsfs}
\usepackage{graphicx,amsthm,enumerate,enumitem}
\usepackage[titletoc,toc]{appendix}
\usepackage{natbib,algorithm2e}
\usepackage{setspace,tabularx,booktabs,adjustbox,calc}
\usepackage{subcaption}
\newcommand{\zap}[1]{}

\newcommand{\mcI}{\mathcal{I}}
\newcommand{\mcW}{\mathcal{W}}
\newcommand{\mcS}{\mathcal{S}}
\newcommand{\Xbar}{\bar{X}}
\newcommand{\NSGS}{NSGS$_p$}
\makeatletter
\newcommand{\nosemic}{\renewcommand{\@endalgocfline}{\relax}}% Drop semi-colon ;
\newcommand{\dosemic}{\renewcommand{\@endalgocfline}{\algocf@endline}}% Reinstate semi-colon ;
\newcommand{\pushline}{\Indp}% Indent
\newcommand{\popline}{\Indm\dosemic}% Undent
\makeatother

\newcolumntype{L}[1]{>{\raggedright\let\newline\\\arraybackslash\hspace{0pt}}m{#1}}
\newcolumntype{C}[1]{>{\centering\let\newline\\\arraybackslash\hspace{0pt}}m{#1}}
\newcolumntype{R}[1]{>{\raggedleft\let\newline\\\arraybackslash\hspace{0pt}}m{#1}}
\newcolumntype{Y}{>{\centering\arraybackslash}X}

\usepackage{natbib}
 \bibpunct[, ]{(}{)}{,}{a}{}{,}%

\newtheorem{theorem}{Theorem}

\newtheorem{lemma}{Lemma}
\newtheorem{assumption}{Assumption}

\begin{document}
\title{Efficient Ranking and Selection in Parallel Computing Environments}
\author{Eric C. Ni\\
    %School of
%    Operations Research and Information Engineering\\
    Cornell University\\
    Ithaca, New York 14853, USA\\ \texttt{cn254@cornell.edu}
    \and
    Dragos F. Ciocan \\
%    Technology and Operations Management\\
    INSEAD\\
    Fontainebleau 77305, FRANCE\\ \texttt{florin.ciocan@insead.edu}
    \and
    Shane G. Henderson\\
%    Operations Research and Information Engineering\\
    Cornell University\\
    Ithaca, New York 14853, USA\\\texttt{sgh9@cornell.edu}
    \and
    Susan R. Hunter \\
%    School of Industrial Engineering \\
    Purdue University\\
    West Lafayette, Indiana 47907, USA\\\texttt{susanhunter@purdue.edu}
}
\date{}
\maketitle
\begin{abstract}
The goal of ranking and selection (R\&S) procedures is to identify the best stochastic system from among a finite set of competing alternatives. Such procedures require constructing estimates of each system's performance, which can be obtained simultaneously by running multiple independent replications on a parallel computing platform. However, nontrivial statistical and implementation issues arise when designing R\&S procedures for a parallel computing environment. Thus we propose several design principles for parallel R\&S procedures that preserve statistical validity and maximize core utilization, especially when large numbers of alternatives or cores are involved. These principles are followed closely by our parallel Good Selection Procedure (GSP), which, under the assumption of normally distributed output, (i) guarantees to select a system in the indifference zone with high probability, (ii) runs efficiently on up to 1,024 parallel cores, and (iii) in an example uses smaller sample sizes compared to existing parallel procedures, particularly for large problems (over $10^6$ alternatives). In our computational study we discuss two methods for implementing GSP on parallel computers, namely the Message-Passing Interface (MPI) and Hadoop MapReduce and show that the latter provides good protection against core failures at the expense of a significant drop in utilization due to periodic unavoidable synchronization.
\end{abstract}

%% \subsection{Keywords for paper submission}
%% Your NIPS paper can be submitted with any of the following keywords (more than one keyword is possible for each paper):

%% \begin{verbatim}
%% Bioinformatics
%% Biological Vision
%% Brain Imaging and Brain Computer Interfacing
%% Clustering
%% Cognitive Science
%% Control and Reinforcement Learning
%% Dimensionality Reduction and Manifolds
%% Feature Selection
%% Gaussian Processes
%% Graphical Models
%% Hardware Technologies
%% Kernels
%% Learning Theory
%% Machine Vision
%% Margins and Boosting
%% Neural Networks
%% Neuroscience
%% Other Algorithms and Architectures
%% Other Applications
%% Semi-supervised Learning
%% Speech and Signal Processing
%% Text and Language Applications

%% \end{verbatim}

\section{Introduction}

The simulation optimization (SO) problem is a nonlinear optimization problem in which the objective function is defined implicitly through a Monte Carlo simulation, and thus can only be observed with error. Such problems are common in a variety of applications including transportation, public health, and supply chain management; for these and other examples, see \texttt{SimOpt.org} \citep{simoptlib}. For overviews of methods to solve the SO problem, see, e.g., \cite{1994fu,1998andWSC,2005fugloaprWSC,2013pasgho}. 

We consider the case of SO on finite sets, in which the decision variables can be  categorical, integer-ordered and finite, or a finite ``grid'' constructed from a continuous space.  Formally, the SO problem on finite sets can be written as   
%
%Simulation optimization (SO) involves solving problems of the form
\begin{align}
\quad \max_{i\in \mcS} \,\, \mu_i=E[X(i;\xi)] \label{eq:problem}
\end{align}
where $\mcS=\{1,\hdots,k\}$ is a finite set of design points or ``systems'' indexed by $i$, and $\xi$ is a random element used to model the stochastic nature of simulation experiments. 
(In the remainder of the paper
we assume %, unbeknownst to the selection procedure, 
that $\mu_1 \le
\mu_2 \le \cdots \le \mu_k$, so that system~$k$ is the best.)  
The objective function $\mu:\mcS\rightarrow\mathbb{R}$ cannot be computed exactly, but can be estimated using output from a stochastic simulation represented by $X(\cdot;\xi)$. While the feasible space $\mcS$ may have topology, as in the finite but integer-ordered case, we consider only methods to solve the SO problem in \eqref{eq:problem} that (i) do not exploit such topology or structural properties of the function,  and that (ii)  apply when the computational budget permits at least \emph{some} simulation of \emph{every} system. Such methods are called \emph{ranking and selection} (R\&S) procedures. 
%Because of the randomness in $X(\cdot;\xi)$ and the potentially large solution
%space $\mcS$, providing a probabilistic guarantee on the quality of the solution found %solving SO problems 
%can be computationally costly. %For the same
%reason, it can be difficult for an SO algorithm to guarantee the
%quality of the solution it finds. 

%For examples of SO problems, see \texttt{SimOpt.org} \citep{simoptlib}. 

%In this paper, we focus on Ranking and Selection (R\&S) 
%R\&S procedures are a  class of SO methods that apply when $\mcS$ is finite %, say $\mcS = \{1,2,\ldots,k\}$, 
%and our computational budget permits at least {\em some} simulation of {\em every} system. 
R\&S procedures are frequently used in simulation studies because %they do not rely on 
structural properties, such as convexity,  are difficult
to verify for simulation models and rarely
hold. They can also be used in conjunction with heuristic search
procedures in a variety of ways
\citep{Pichitlamken2006,BNK2003}, making them useful even if not all
systems can be simulated. See \cite{KN2006} for an excellent
introduction to, and overview of, R\&S procedures. We remark here that while R\&S problems are
closely related to best-arm problems, there are several differences between these bodies of literature. Almost always, %Typically, although 
the algorithms developed
in the best-arm literature %almost always 
assume that only one system
is simulated at a time %; see, e.g.,
\citep[see, e.g.,][]{Jamieson2014,bubces12} and that %. Moreover, best-arm problems almost always assume that 
simulation outputs are bounded, or that all
variances have a known bound. %are bounded, and the bounds are known.

%\subsection{Parallel Ranking and Selection}
%R\&S procedures typically demand multiple independent replications across many systems. 
%In classical R\&S literature, the search for optimal solution 
% simulation output are generated sequentially, following
%specific procedures 
%that 
%determine which systems to simulate and their corresponding sample sizes. After each new sample is observed, there is usually a ``screening'' operation which calculates sample statistics and eliminates inferior systems. This process is iterated with new samples obtained and screenings performed after each iteration, until a termination criteria is met. R\&S procedures are 
%is essentially made up of three computational tasks: (1) deciding what simulations to run next, (2) running simulation, and (3) screening (computing statistical estimators and determining which systems are inferior). These tasks are performed in specific orders, often repeatedly, until a termination criteria is met. 

R\&S procedures are  designed to offer one of several types of probabilistic guarantees, and can be Bayesian or frequentist in nature. 
Bayesian procedures offer guarantees related to a loss function associated with a non-optimal choice; see \cite{BCS2007} and \citet[][Chapter~3]{Chen2015}. 
Frequentist procedures typically offer one of two statistical guarantees; in defining these guarantees, let $\delta>0$ be a known constant and let $\alpha\in(0,1)$ be a parameter selected by the user. 
The {\em  Probability of Correct Selection} (PCS) guarantee is a guarantee that, whenever $\mu_k -\mu_{k-1} \ge \delta$, % for some known $\delta > 0$, 
the probability of selecting the best system $k$ when the procedure terminates is greater than $1-\alpha$. %, where $\alpha\in (0,1)$ is chosen by the user. 
Henceforth, the assumption that $\mu_k -\mu_{k-1} \ge \delta$ will be called the  {\em PCS assumption};  
if $\mu_k - \mu_{k-1} < \delta$ then a PCS guarantee does not hold.
In contrast, the {\em Probability of Good Selection} (PGS) guarantee is a guarantee that the probability of selecting a system with objective value 
within $\delta$ of the best %, again with $\delta > 0$ known, 
is greater than $1-\alpha$. % , where $\alpha\in(0,1)$ is chosen by the user. 
That is, the PGS guarantee implies 
PGS $= \mathrm{P}[\text{Select a system } K \text{ such that } {\mu_k-
    \mu_K\le \delta}] \geq 1-\alpha$. 
A PGS guarantee makes no assumption about the configuration of the means and %. % vector $\mu$.
%Most R\&S procedures are designed to guarantee a PCS or PGS of at least $1-\alpha$, where $\alpha\in (0,1)$ is chosen by the user. 
%A PGS guarantee 
is the same as the ``probably approximately correct'' guarantee in best-arm literature.

Traditionally, R\&S procedures were limited to problems with a modest
number of systems $k$, say $k \le 100$, due to the need to assume
worst-case mean configurations to construct validity proofs. The
advent of screening, i.e., discarding clearly inferior alternatives
early on \citep{Nelson2001,Kim2006,Hong2006}, has allowed R\&S to be
applied to larger problems, say $k \le 500$. Exploiting parallel
computing is a natural next step as argued in, e.g., \cite{fu02}. By employing  parallel cores, simulation output can be generated at a higher rate, and a parallel R\&S 
procedure should complete in a smaller amount of time than its
sequential equivalent, and allowing larger problems to be solved.

%\susan{Consider an R\&S procedure that is designed to solve the SO problem in \eqref{eq:problem}. Such a  procedure will estimate the performances of the systems by obtaining multiple simulation replications from each system. Then, upon some termination criteria, the algorithm will return to the user the estimated-best system. Crucial in this step is obtaining \emph{multiple simulation replications} of each system, which makes the SO problem in \eqref{eq:problem} an excellent candidate for implementation on parallel computing platforms. Thus many cores can be used to generate independent replications concurrently.  By employing more parallel cores, simulation output can be generated at a higher rate, and a parallel R\&S  procedure should complete in a smaller amount of time than its sequential equivalent. }

%Therefore, if the number of systems in the R\&S problem is large, one should expect a large amount of computation before a procedure terminates. Such demanding computational cost required by R\&S procedures has limited the usefulness for large-scale problems.

%Even if the high computational cost associated with large-scale R\&S problems seems inexorable, using parallel computing can clearly make generating simulation replications, hence solving those problems, more quickly. 

\cite{hei88,glyhei90,glyhei91} explored the use of parallel computers
to construct valid simulation estimators, but R\&S procedures that
exploit parallel computing have emerged only recently. \cite{Luo2000}
and \cite{Yoo2009} employ a web-based computing environment and
present a parallel procedure under the optimal computing budget
allocation (OCBA) framework. (OCBA has impressive empirical
performance, but does not offer PCS or PGS guarantees.)
\cite{Chen2005} tests a sequential pairwise hypothesis testing
approach on a local network of computers. More recently,
%\cite{Luo2011} tackled a problem with the number of alternatives on the order of 3000, and  \
\cite{Luo2013} develop a parallel adaptation of a
fully-sequential R\&S procedure that provides an asymptotic (as
$\delta \to 0$) PCS guarantee. %\cite{Ni2014} present a parallelization of the two-stage NSGS \citep{Nelson2001} procedure. 
\cite{Luo2013} is the best known existing method for parallel ranking and selection that provides a form of PCS guarantee on the returned solution.

In this paper, we (i) identify opportunities and challenges that arise
from adopting a parallel computing environment to solve large-scale
R\&S problems, (ii) propose a Good Selection Procedure (GSP) that solves R\&S problems on parallel computers, and (iii) implement our procedure in two different parallel computing frameworks. We make the following contributions.

%\eric{CONTRIBUTIONS VERSION 1}
\textbf{Theoretical contributions.}  We propose a number of design principles that promote
efficiency and validity in such an environment, and demonstrate them
in a new parallel GSP. % Good Selection Procedure (GSP). 
GSP showcases the power of these design principles in that it greatly
extends the boundary on the size of solvable R\&S problems. While the
%best known existing 
method of \citep{Luo2013} can solve on the order of
$10^4$ systems, one of our implementations of GSP is capable of
solving R\&S problems with more than $10^6$ systems. Our computational
results include such a problem, which we solve in under 6 minutes
on $10^3$ cores.
% (needs to be one paragraph, otherwise, should use section header, list, or something else to distinguish from the surrounding text.)
Another important theoretical contribution of this paper is the redesigned
screening method in GSP which, unlike many fully-sequential
procedures \citep{KimNels01,Hong2006}, does not rely on the PCS
assumption. Accordingly, many systems can lie within the
indifference-zone, i.e., have an objective function value within
$\delta$ of that of System $k$, as will usually be the case when the
number of systems is very large. Our procedure then provides the same PGS
guarantee as existing indifference-zone procedures like
\cite{Nelson2001} but with far smaller sample sizes.

\textbf{Practical contributions.}
The GSP procedure discussed in this paper is intended for any parallel, shared or non-shared memory platform where cores can communicate with each other. As long as no core fails during execution, it should deliver expected results regardless of the hardware specification. 
%In reality, as computer hardware differs significantly in communication speed, data storage, reliability, and cost structure, the actual execution speed and cost-effectiveness of a parallel algorithm implementation clearly depends on how well the it uses the hardware. 
%For instance, holding other things equal, communication speed is negatively correlated with the physical distance between cores, and can be orders of magnitude faster on a local cluster compared to a distributed computer cloud. As a consequence, procedures that require intensive message-passing would therefore be much less favorable on the latter architecture.
%There are various parallel programming software tools that engage the hardware in very different ways. 
The procedure is also amenable to a range of existing parallel computing frameworks. We offer implementations of GSP based on MPI (Message-Passing Interface) and Hadoop MapReduce, and show how they differ in construction and in performance. The reasons for our choice of implementation frameworks are twofold:
\begin{itemize}
    \item Both MPI and MapReduce are among the most popular and mature platforms for deploying parallel code, on a wide range of systems ranging from high performance supercomputers to commodity clusters such as Amazon EC2.
    \item MPI and MapReduce provide points of comparison between two different parallel design philosophies. Broadly speaking, the former enables low level tailoring and optimization in the implementation of a parallel procedure, while the latter is more of a ``one-size-fits-all'' framework which delegates as  much of the implementation complexity as possible to the MapReduce package itself. 
\end{itemize}
As we shall see, MPI is the more efficient of the two, achieving speed and utilization gains of around a factor of magnitude over MapReduce. On the other hand, MapReduce offers acceptable performance for large scale problems, and is more robust to reliability issues that may arise in cloud-computing environments where parallel tasks may fail to complete due to unresponsive cores.

The remainder of the paper is organized as
follows. \S\ref{sect:DesignPrinciples} discusses the design
principles followed in creating GSP to promote efficiency and ensure
the procedure's validity. \S\ref{sect:GoodSelectionProcedure}
describes our multi-stage parallel R\&S procedure GSP, and establishes
the PGS guarantee. Computational studies in
\S\ref{sect:computation} support our assertions on the quality
of GSP and its parallel implementations, and point to open-access
repositories where the code can be obtained. An appendix
contains more proof detail, and further information on the MPI and
MapReduce implementations.
This paper is a considerable outgrowth of the conference papers \cite{Ni2013,Ni2014,Ni2015}.

\section{Design Principles for Parallel R\&S Procedures}%==================================================
\label{sect:DesignPrinciples}

R\&S procedures are essentially made up of three computational tasks:
(1) deciding what simulations to run next, (2) running simulations,
and (3) screening (computing statistical estimators and determining
which systems are inferior). On a single-core computer, these tasks
are repeatedly performed in a certain order until a termination criterion is met. On a parallel platform, multiple cores can simultaneously perform one or several of these tasks.

In this section, we discuss various issues that arise when a R\&S procedure is designed for and implemented on parallel platforms to solve large-scale R\&S problems. We argue that failing to consider these issues may result in impractically expensive or invalid procedures. We recommend strategies by which these issues can be addressed, and illustrate how we incorporate them in our procedure presented in \S\ref{sect:GoodSelectionProcedure}, which iteratively runs Tasks~(1) through (3) in multiple stages.
%iterative screening to eliminate clearly inferior systems, and incorporates a \cite{Rinott1978} step to select the best system from among the good ones. 

For discussing the design principles for parallel R\&S procedures in this section, we consider a parallel computing environment that satisfies the following properties. 
\begin{assumption}\label{assum:Independence}
    (\textit{Core Independence}) A fixed number of processing units (``cores'') are employed to execute the parallel procedure. Each core is capable of performing its own set of computations without interfering with other cores unless instructed to do so. Each core has its own memory and does not access the memory of other cores.
\end{assumption}
\begin{assumption}\label{assum:Message-passing}
    (\textit{Message-passing}) The cores are capable of communicating through sending and receiving messages of common data types and arbitrary lengths.
\end{assumption}
\begin{assumption}\label{assum:Reliability}
    (\textit{Reliability}) Cores do not ``fail'' or suddenly become unavailable. Messages are never ``lost''.
\end{assumption}
Many parallel computer platforms satisfy the first two assumptions,
but some are subject to the risk of core failure, which may interrupt
the computation in various ways. For clarity, we work under the
reliability assumption and defer the design of failure-proof
procedures to  \S\ref{sect:implementation_hadoop} where we
discuss Hadoop MapReduce.
Similar to \cite{Luo2013} and \cite{Ni2013}, we consider a
master-worker framework, using a uniquely executed ``master'' process
(typically run on a dedicated ``master'' core) to coordinate the
parallel procedure, and letting other cores (the ``workers'') work
according to the master's instructions. To the extent possible we want
to avoid synchronization delays, where one core cannot continue until
another core completes its task, as we will see in \S\ref{sect:implementation_parallel}.

\subsection{Implications of Random Completion Times} \label{sect:RandCompTime}

Consider the simplest case where only Task~(2), running simulations, is run in parallel, and each simulation replication completes in a random amount of time. To construct estimators for a single system simulated by multiple cores, one can either collect  a fixed number of replications in a random completion time, or a random number of
replications in a fixed completion time \citep{hei88}. \cite{hei88} and
\cite{glyhei90,glyhei91} discuss unbiased estimators of each
type. 
%There are two types of estimators proposed by \cite{hei88} to collect replications of a single system conducted across multiple cores: 
%(i)
%estimators that produce a fixed number of replications in a random
%completion time, and 
%(ii) estimators that produce a random number of replications in a fixed completion time. \cite{hei88} and
%\cite{glyhei90,glyhei91} discuss unbiased estimators of each type. 
%Type~(ii) estimators are more difficult to incorporate in our context than type~(i) estimators 
%because a random number of replications collected after a fixed amount of time may not be i.i.d. with the desired distribution upon which much of the screening
%theory depends \citep{hei88,glyhei91,Ni2013,Luo2013}. 
Because a random number of replications collected after a fixed amount of time may not be i.i.d. with the desired distribution upon which much of the screening
theory depends \citep{hei88,glyhei91,Ni2013,Luo2013}, 
%In general, screening in the setting of a random number of completed replications requires great care and perhaps new theory. 
%Thus 
we confine our attention to estimators that produce a fixed number of replications in a random completion time. (The cause of
this difficulty can be traced to dependence between the estimated
objective function and computational time.)

Using estimators that produce a fixed number of replications in a random completion time %this type of estimators 
for parallel R\&S places a restriction on the manner in which replications can validly be  farmed out to and collected from the workers. Consider the case where more than one core simulates the same system, and replications generated in parallel are aggregated to produce a single estimator. A na\"{i}ve way is to collect replications from any core following the order in which they are generated, but as demonstrated in \citet[\S3.1]{Ni2013}, the estimators may be biased, making it hard to establish provable statistical guarantees. In contrast, a valid method is to place the finished replications in a predetermined order and use them as if they are generated following that order, to avoid ``re-ordering'' of the simulation replications caused by random completion time. 

Under this principle, our GSP  in \S\ref{sect:GoodSelectionProcedure} ensures that the simulation results generated in parallel are initiated, collected, assembled and used by the screening routine in an ordered manner. Specifically, in Stage~2 of GSP, when the master instructs a worker to simulate system~$i$ for a batch of replications (Step~\ref{step:Stage2Main}\ref{step:NHHsim}), the batch index is also received by the worker. When the batch is completed, its statistics are sent back to the master alongside the batch index (Step~\ref{step:Stage2Main}\ref{step:NHHsim}), which signals its pre-determined position in the assembled batch sequence on the master. This ensures that the batch statistics sent to workers for screening (Step~\ref{step:Stage2Main}\ref{step:NHHscreening}) follow the exact order in which they were initiated, and constructed estimators are unbiased with the correct distribution. \cite{Luo2013} discuss a similar approach which they refer to as ``vector-filling''. %{If page count is an issue, the description of how batch indices are handled could be shortened.}

%With the ability to construct unbiased, i.i.d. samples from simulation replications generated in parallel, we are able to propose a simple parallel R\&S model as follows. Let simulation replications be generated in parallel across multiple cores, assembled in a predetermined order as mentioned above, and passed to another core running some sequential R\&S procedure. The only difference of this R\&S model compared to a purely sequential procedure is that simulation replications are generated in parallel instead of on a single core. Since the assembled replications still provide i.i.d. samples from the same distribution, the new procedure should offer the same statistical validity. In the following subsection, we will discuss the limitations of this simple model in a realistic parallel computing environment and our solutions.

%There exist other ways to construct estimators that can be used by R\&S procedures facing random simulation completion times. For instance, \cite{Luo2013} proposed an ``asymptotically valid'' procedure, using estimators whose bias diminish in a particular asymptotic regime. 

\subsection{Allocating Tasks to the Master and Workers}
%Recall that R\&S procedures are made up of three tasks: (1) deciding what simulations to run next, (2) running simulations, and (3) screening.
Previous work on parallel R\&S procedures \citep{Chen2000,Yoo2009,Luo2011,Luo2013} focuses almost exclusively on pushing Task~(2), running simulations, to parallel cores. In those procedures, usually the master is solely responsible for Tasks~(1) and (3), deciding what simulations to run next and screening, and the workers perform Task (2) in parallel.
%Our simple parallel R\&S model in 
%the previous subsection 
%Section~\ref{sect:RandCompTime}
%also had (2) done in parallel but keeps (1) and (3) on a single master. 
In this setting, the benefit of using a parallel computing platform is entirely attributed to distributing simulation across parallel cores, hence reducing the total amount of time required by Task~(2).

However, the master could potentially become a bottleneck in a number of ways. 
First, as noted by \cite{Luo2011}, the master can be overwhelmed with messages.
Second, for the master to keep track of all simulation results requires a large amount of memory, especially when the number of systems is large \citep{Luo2013}. 
Finally, when the number of systems is large and simulation output is generated by many workers concurrently, running Tasks~(1) and (3) on the master alone may become relatively slow, resulting in a waste of core hours on workers waiting for the master's further instructions.
Therefore, a truly scalable parallel R\&S procedure should allow its users a simple way to control the level of communication, use the memory efficiently, and distribute 
as many tasks as possible
%screening as well as simulation 
across parallel cores. In addition, it should perform some form of load-balancing to minimize idling on workers.

\subsubsection{Batching to Reduce Communication Load} \label{sect:batching}
One way to reduce the number of messages handled by the master is to control communication frequency by having the workers run simulation replications in batches and only communicate once after each batch is finished. 

%{
Since R\&S procedures typically use summary statistics rather than individual observations when screening systems, it may even suffice for the worker to compute and report batch statistics instead of point observations from every single replication. Indeed, a useful property of our statistic for screening systems $i$ and $j$ is that it is updated using only the sample means over the entirety of the most recent batch $r$, instead of requiring the collection of individual replication outcomes. These sample means can be independently computed on the worker(s) running the $r$th batch of systems $i$ and $j$, and the amount of communication needed in reporting them to the master is constant and does not grow with the batch size.

The distribution of batches in parallel must be handled with care. Most importantly, since using a random number of replications after a fixed run time may introduce bias (as we have shown in \S\ref{sect:RandCompTime}), a valid procedure should employ a predetermined and fixed batch size for each system, which may vary across different systems.
%, so long as the batch sizes are independent of completion time. 
Batches generated in parallel for the same system should be assembled according to a predetermined order, following the same argument used in \S\ref{sect:RandCompTime}. Furthermore, if the procedure requires screening upon completion of every batch, then it is necessary to perform screening steps following the assembled order.

\subsubsection{Allocating Simulation Time to Systems} \label{sect:BatchSizes}

When multiple systems survive a round of screening, R\&S procedures need to decide which system(s) to simulate next (possibly on multiple cores), and how many replications to take. 
%Sequential R\&S procedures make the first decision in one of two ways. \textit{Adaptive} sampling \citep{Hong2006,HN2005} considers past samples and selects one system to simulate, whereas \textit{grouped} sampling \citep{Nelson2001,KimNels01,Kim2006} involves generating a sample from each surviving system before the next round of screening.
While sequential procedures usually sample one replication from the chosen system(s), or multiple replications from a single system, it is natural for a parallel procedure to consider strategies that sample multiple replications from multiple systems. 
%It remains to choose the sample size of each surviving system.
%which, as we have discussed in Section \ref{sect:batching}, is required to be fixed independent of .
In doing so, the parallel procedure may adopt sampling strategies such that simulation resources are allocated to surviving systems in a most efficient manner. 

The best practice in making such allocations depends on the specific
screening method. For instance, in \cite{Hong2006} as well as GSP,
screening between systems $i$ and $j$ is based on a scaled Brownian
motion $B([\sigma_i^2/n_i+\sigma_{j}^2/n_{j}]^{-1})$ where $B(\cdot)$
denotes a standard Brownian motion (with zero drift and unit volatility), $n_i$ is the sample size and
$\sigma_i^2$ is the variance of system $i$. To drive this Brownian
motion rapidly with the fewest samples possible, which accelerates
screening, \cite{Hong2006}
recommended that the ratio $n_i/\sigma_i$ be kept equal across all
surviving systems. 

The above recommendation implicitly assumes that simulation completion time is fixed for all systems, and is suboptimal when completion time varies across systems. Suppose all workers are identical, and each replication of system $i$ takes a fixed amount of time $T_i$ to simulate on any worker. We can then formulate the problem of advancing the above Brownian motion as 
\begin{align*}
\max\,\,\, &[\sigma_i^2/n_i+\sigma_{j}^2/n_{j}]^{-1}\\
\text{s.t. } & n_iT_i+n_{j}T_{j}=T
\end{align*}
which yields the optimal computing time allocation 
\begin{align}
\frac{n_iT_i}{n_{j}T_{j}}=\frac{\sigma_i\sqrt{T_i}}{\sigma_{j}\sqrt{T_{j}}}. 
\label{eq:optimBatchSize}
\end{align}
This result is consistent with a conclusion in \cite{glywhi92}, that when simulation completion time $T_i$ varies, an asymptotic measure of efficiency per replication is inversely proportional to $\sigma_i^2 E[T_i]$.
%, which is consistent with the above result.

In practice, $T_i$ is unknown and possibly random, so both $E[T_i]$ and $\sigma^2$ need to be estimated in a preliminary stage. Suppose they are estimated by some estimators $\bar{T_i}$ and $S_i^2$. Then we recommend setting the batch size for each system $i$ proportional to $S_i/\sqrt{\bar{T_i}}$ following \eqref{eq:optimBatchSize}.
%Note that conditioning 
%As $T_i$ and $X_i$ may be correlated, in our procedure, we estimate $T_i$ using an independent sample in Stage~0 and do not include the simulation output from Stage~0 in screening statistics. On the other hand, since the Stage~1 sample mean $\bar{X}_i(n_1)$ and variance $S_i^2$ are independent, the Stage~1 sample mean can still be used for Stage~2 screening without bias. In general, batch sizes may depend on simulation output, but such dependency should not invalidate the normality and unbiasedness required of screening statistics.

%Two types of bottlenecks: Bottlenecks of processing too many messages, and bottlenecks of too many ``screening'' operations. 
%\begin{small}
%
%
%\begin{theorem}[in pencil]
%The approximate ``batch size'' should be $(?)$ as a function of the number of systems, the number of cores, and the time to complete a replication of each system (may vary by system).
%\end{theorem}
%\begin{proof}
%Consult Hong and Nelson ``sampling versus switching" paper for insight on this result.
%\end{proof}
%
%\begin{itemize*}
%\item To reduce computational burden of screening, we screen at the workers in the dedicated phase, and screen at the master in the re-distributed phase. 
%\item (Can we provide any idea of how many cores would be required before screening and/or zipping at the master is a bottleneck, even in the re-distributed phase? Actually, batching helps with this.)
%\end{itemize*}
%\end{small}

\subsubsection{Distributed screening} \label{sect:DistributedScreening}
In fully sequential R\&S procedures, e.g., \cite{Kim2006,Hong2006},
each screening step typically involves doing a fixed amount of
calculation between every pair of systems to decide if one system is
better than another with a certain degree of statistical
confidence. The amount of work is proportional to the number of pairs
of systems, which is $O(k^2)$.

In the serial R\&S literature, the computational cost of screening is assumed to be negligible compared to that of simulation because the number of systems $k$ is usually quite small
%.
%because 
%As many as $k$ replications are taken sequentially before each screening is performed, 
and each simulation replication may take orders of magnitude longer than $O(k^2)$ screening operations required in each iteration. Under this assumption, it is tempting to simply have the master handle all screening after the workers complete a simulation batch. This approach can easily be implemented and proven to be statistically valid. However, it may become computationally inefficient because all workers stay idle while the master screens, so a total amount of $O(ck^2)$ processing time is wasted, where $c$ is the number of workers. For a large problem with a million systems solved on a thousand cores, the wasted processing time per round of screening can easily amount to thousands of core hours, reducing the benefits from a parallel implementation dramatically. Moreover, if the procedure requires computing and storing in memory some quantities for each system pair (for instance, the variance of differences between systems), then the total amount of $O(k^2)$ memory may easily exceed the limit for a single core.

%\begin{figure}
%\centering
%\begin{subfigure}[b]{0.5\textwidth}
%\includegraphics[width=\textwidth]{figures/pairwise_screen2.eps}
%\caption{Screening on the Master}
%\label{fig:screenOnMaster}
%\end{subfigure}%
%\begin{subfigure}[b]{0.5\textwidth}
%\includegraphics[width=\textwidth]{figures/pairwise_screen1.eps}
%\caption{Screening on the Workers}
%\label{fig:screenOnWorkers}
%\end{subfigure}
%\caption{Comparisons between different screening methods. Each dot represents a pair of systems to be screened against each other.}\label{fig:distributedScreening}
%\end{figure}
\begin{figure}
    %\FIGURE
    %{
    %\includegraphics*[width=0.5\textwidth]{figures/pairwise_screen2.eps}}
    %{Screening on the Master. \label{fig:screenOnMaster}}{}
    \centering
    {
        \includegraphics*[width=0.4\textwidth]{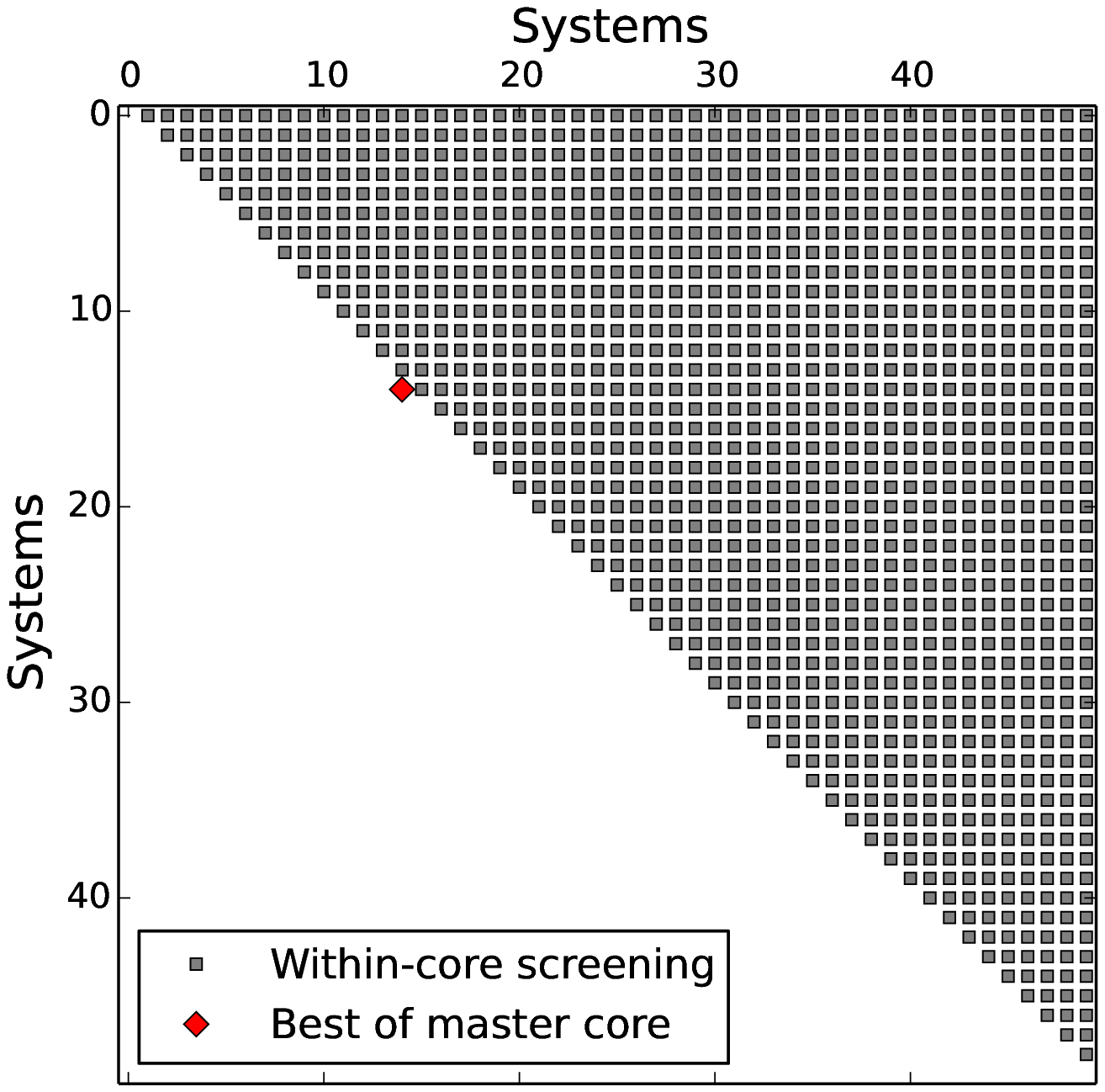}
        \includegraphics*[width=0.4\textwidth]{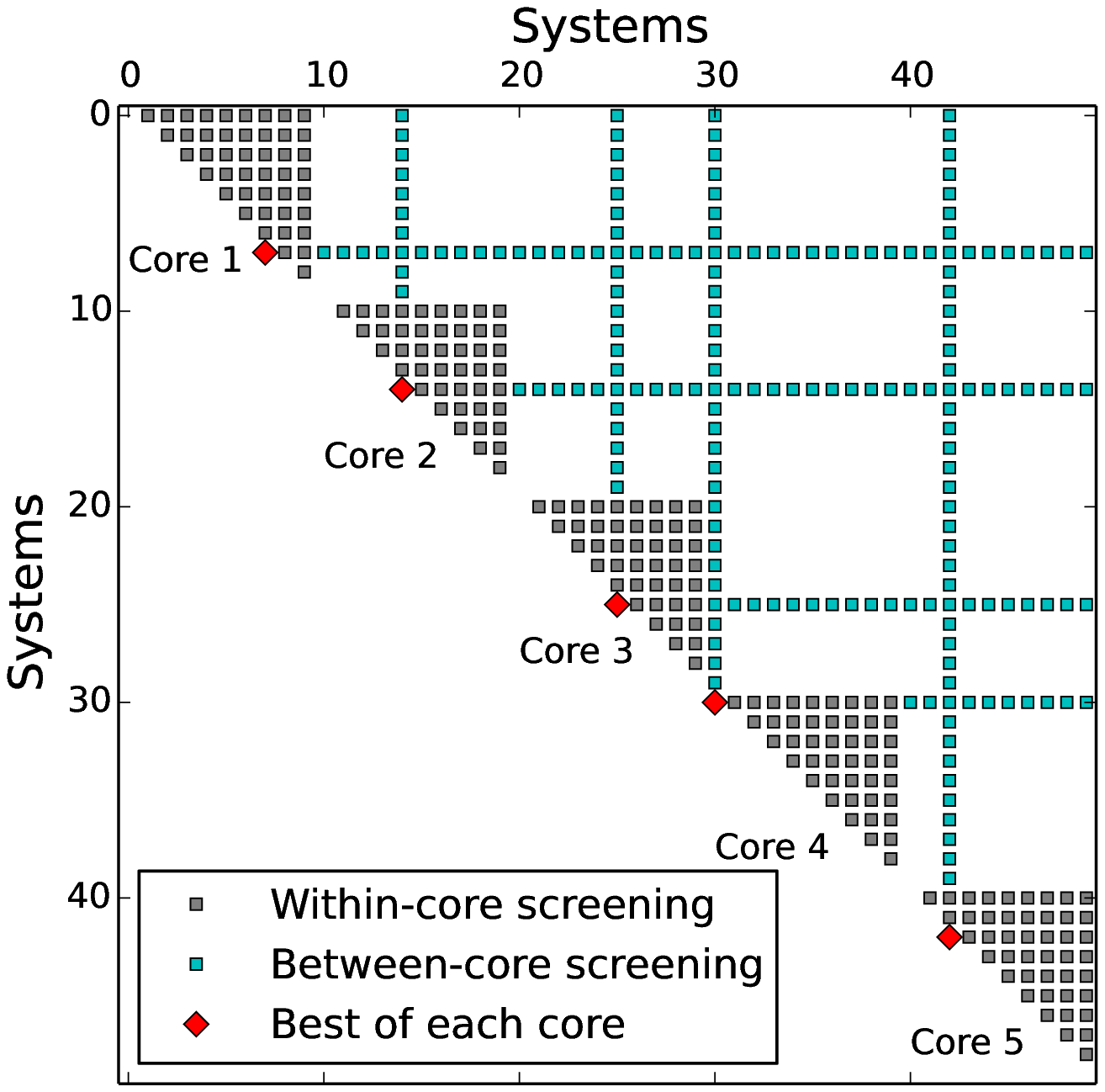}
    }
    \caption{Comparison of screening methods applied on 50 systems. Each black or green dot represents a pair of systems to be screened. In the left panel, all pairs of screening is done on the master. In the right panel, each worker core gets 10 systems, screens between themselves, and screens its systems against one system from every other worker that has the highest sample mean.  \label{fig:screenOnWorkers}}
    %{More explanations here.}
    {}
\end{figure}
It is therefore worth considering strategies that distribute screening among workers. A natural strategy is to assign roughly $k/c$ systems to each worker, and let it screen among those systems only, as illustrated in Figure~\ref{fig:screenOnWorkers}. By doing so, each worker screens $k/c$ systems, occupying only $O(k^2/c^2)$ memory, and performing $O(k^2/c^2)$ work in parallel. Hence the wall-clock time for each round of screening is reduced by a factor of $c^2$.

Under the distributed screening scheme, not all pairs of systems are
compared, so fewer systems may get eliminated. The reduction in
effectiveness of screening can be compensated by sharing some good
systems across workers. In Figure~\ref{fig:screenOnWorkers}, for
example, each core shares its own (estimated) best system with other
cores, and each system is screened against other systems on the same
core, as well as $O(c)$ good systems from other cores. This greatly
improves the chance that each system is screened against a good one,
despite the extra work to share those good systems. As illustrated in
Figure~\ref{fig:screenOnWorkers}, the additional number of
pairs that need to be screened on each core is only
$O(k)$ when the best system on each core is shared. Alternatively, the
procedure may also choose to share only a smaller number $c'\ll c$ of
good systems, so that the communication workload associated with this
sharing does not increase as the number of workers increases. 

%\eric{Remove this paragraph? our procedure is now based on a ``there exists a path'' argument and catch-up screening is not necessary.}
The statistical validity of some screening-based R\&S procedures
(e.g. \citealt{KimNels01,Hong2006,Luo2013}) requires screening to be
performed once every replication (or batch of replications) is
simulated. This implies that, when the identity of the
estimated-best system(s) changes, the master has to communicate all
previous replication results of the new estimated-best system(s) to
the workers, so that they can perform all of the screening steps up to
the current replication to ensure validity of the screening. (If
screening on a strict subsequence of replications, it may be
sufficient to communicate summary statistics.) Such ``catch-up''
screening was used, for instance, in \cite{Pichitlamken2006}, in a
different context. In \S\ref{sect:GoodSelectionProcedure}, we
employ a probabilistic bound that removes the need for catch-up
screening in GSP.

Besides core hours, distributing screening across workers also saves memory space on the master. In our implementation of GSP, the master keeps a complete copy of batch statistics only for a small number of systems that are estimated to be the best. For a system that is not among the best, the master acts as an intermediary, keeping statistics for only the most recent batches that have not been collected by a worker. Whenever some batch statistics are sent to a worker (for screening in Steps~\ref{step:Stage1Main}\ref{step:NHHStage1Screening} or \ref{step:Stage2Main}\ref{step:NHHscreening} of GSP), they can be deleted on the master. This helps to even out memory usage across cores, making the procedure capable of solving larger problems without the need to use slower forms of storage.

\subsection{Random Number Stream Management}
The validity and performance of simulation experiments and simulation optimization procedures relies substantially on the quality and efficiency of (pseudo) random number generators. 
%It is a common practice to use a deterministic algorithm to generate a sequence of numbers in $[0,1]$ which resembles a truly random sample from the Uniform$(0,1)$ distribution. 
For a discussion of random number generators and their desirable properties, see \cite{LEcuyer2006}.

%Random number streams are widely used for managing independence in random number generation. \cite{Pasupathy2011} recommended On a single-core computer, one typically need to ensure independence of simulation output between system and
%($Y_{i0k}\perp Y_{j0k}$, $i\ne j$, for any $k$), 
%between different replications of the same system. It is also recommended to use 
% ($Y_{i0k}\perp Y_{i0k'}$, $k\ne k'$, for any $i$), 

%On a distributed-memory platform, simulation output are generated in parallel. 
To avoid unnecessary synchronization, each core may run its own random number generator independently of other cores. Some strategies for generating independent random numbers in parallel have been proposed in the literature. \cite{Mascagni2000} consider a class of random number generators which are parametrized so that each valid parametrization is assigned to one core. \cite{Karl2014} adopt \cite{LEcuyer2002}'s \texttt{RngStream} package, which supports streams and substreams, and demonstrated a way to distribute \texttt{RngStream} objects across parallel cores. 

Both methods set up parallel random number generation in such a way that once initialized, each core will be able to generate a unique, statistically independent stream of pseudo random numbers, which we denote as $U_w$, for each $w=1,2,\ldots, c $. If a core has to switch between systems to simulate, one can partition $U_w$ into substreams $\{U_w^i:i=1,2,\ldots, k\}$, simulating system $i$ using $U_w^i$ only. It follows that for any system~$i$, $U_w^i$ for different $w$ are independent as they are substreams of independent $U_w$'s, so simulation replicates generated in parallel with  $\{U_w^i:w=1,2,\ldots,c\}$ are also i.i.d.
%and derive estimators as if the output $\{X_{iw\ell}: 1\le w\le c , \ell\ge 1\}$ is generated sequentially. 
Moreover, if it is desirable to separate sources of randomness in a simulation, it may help to further divide $U_w^i$ into subsubstreams, each used by a single source of randomness. 

In practice, one does not need to pre-compute and store all random numbers in a (sub)stream, as long as jumping ahead to the next (sub)stream and switching between different (sub)streams are fast. Such operations are easily achievable in constant computational cost; see \cite{LEcuyer2002} for an example.

Although our procedure does not support the use of common random
numbers (CRN), it is worth noting that the above framework easily
extends to accommodate CRN as follows. Begin by having one identical
stream $U_0$ set up on all cores and partitioning it into substreams
$\{U_0(\ell): 1\le \ell\le L\}$ for sufficiently large $L$. Let the
master keep variables $\{\ell_i:i=1,2,\ldots,k\}$  which count the
total number of replications already generated for system~$i$ over all
workers. Each time the master initiates a new replication of system~$i$ on a
worker, it instructs the worker to simulate system~$i$ using substream
$\{U_0(\ell_i+1)\}$ and adds 1 to $\ell_i$. This ensures that for any
$\ell>0$, the $\ell$th replication of every system is generated by the
same substream $\{U_0(\ell)\}$. % {If we need to cut down the page count, maybe the specifics of the extension can be moved to the appendix?}

\section{The Parallel Good Selection Procedure} \label{sect:GoodSelectionProcedure}

In this section, we provide a R\&S procedure GSP that incorporates the design principles from \S\ref{sect:DesignPrinciples}, and is implementable on a wide spectrum of parallel platforms. 
Our procedure applies to the general case in which the system mean and
variance are both unknown and need to be estimated (an earlier version
of the procedure under the known variance case is discussed in
\citealt{Ni2014}), and does not permit the use of common random
numbers. We prove that the procedure offers a PGS guarantee for
normally distributed observations.

\subsection{The Setup}

%Consider the most general case where any system~$i$, $1\le i\le k$, can be simulated on any parallel core~$w$, $1\le w\le c $. We denote the output of the $\ell$th replication of simulating system~$i$ on core~$w$ as a random variable $X_{iw\ell}$ and make the following assumption on its distribution.
%\begin{assumption}\label{assum:mvn}
%For each system $i=1,2,\ldots,k$, the simulation output $\{X_{iw\ell}, {1 \le w \le c }, \ell \ge 1\}$ are i.i.d. replicates from a normal distribution $X_{i}$ with finite mean $\mu_i$ and finite variance $\sigma^2_i$.
%\end{assumption}
%On a single-core computer, $ c =1$ and Assumption \ref{assum:mvn} is typical for many sequential R\&S procedures. In a parallel computing environment, Assumption \ref{assum:mvn} further grants that the distribution of the output random variables or their estimators is solely determined by the system and not affected by the specific core where it is simulated. 
GSP consists of four broad stages.
In an optional Stage~0, workers run $n_0$ simulation replications for each system in parallel to estimate completion times, which are subsequently used to try to balance the workload. As discussed in \S\ref{sect:BatchSizes}, Stage~0 samples are then dropped and not used to form estimators of $\mu_i$'s due to the potential correlation between simulation output and completion time. In Stage~1, a new sample of size $n_1$ is collected from each system to obtain variance estimates $S_i^2=\sum_{\ell=1}^{n_1}(X_{i\ell}-\Xbar_i(n_1) )^2/(n_1-1)$, where $\Xbar_i(n)=\sum_{l=1}^n X_{il}/n$. Prior to Stage~2, obviously inferior systems are screened. In Stage 2, the workers iteratively visit the remaining systems and run additional replications, exchange statistics and independently perform screening over a subset of systems until either all but one are eliminated, or a pre-specified limit on sample size is reached. The screening rule and the limit on sample size are jointly chosen such that inferior systems can be eliminated efficiently, while the best system $k$ survives this stage with high probability regardless of the configuration of true means $\mu_1,\ldots,\mu_k$. Finally, in Stage~3, all systems surviving Stage~2 enter a \cite{Rinott1978} procedure where a maximum sample size is calculated, additional replications are simulated if necessary, and the system with the highest sample mean is selected as the best.

%For clarity, we drop the subscript $j$ and let $X_{i\ell} \sim \text{Normal}(\mu_i,\sigma_i^2)$ denote the outcome of the $\ell$th overall replication for system $i$ collected from Stage~1 onwards. 
%We require that when forming the $\{X_{i\ell}: \ell=1,2,\ldots \}$ sequence using simulation replications from multiple workers, the replications must be sorted in the order in which they are initiated, not when they are completed (more on this in Section~\ref{sect:RandCompTime}). 

The sampling rules used in Stages~0, 1, and 3 are relatively straight
forward, for they each require a fixed number of replications from
each system. In Stage~2, where the procedure iteratively switches
between simulation and screening, a sampling rule needs to be
specified to fix the number of additional replications to take from
each system before each round of screening. Prior to the start of the
overall selection procedure we define increasing (in $r$)
sequences $\{n_i(r): i=1,2,\ldots,k, r=0,1,\ldots \}$ giving the total
number of replications to be collected for system~$i$ by batch $r$,
and let $n_i(0)=n_1$ since we include the Stage~1 sample in mean
estimation. Following the discussion in \S\ref{sect:BatchSizes} where we recommend that batch size for system $i$ be proportional to $S_i/\sqrt{\bar{T_i}}$ in order to efficiently allocate simulation budget across systems, we use
\begin{align}
n_i(r)=n_1+r\left\lceil\beta 
\left(
\frac{S_i}{\sqrt{\bar{T_i}}}
\right)
/
\left(
\frac{1}{k}\sum_{j=1}^{k}
\frac{S_j}{\sqrt{\bar{T_j}}}
\right)
\right\rceil \label{eq:sampling_rule}
\end{align}
where $\bar{T_i}$ is an estimator for simulation completion time of
system~$i$ obtained in Stage~0 if available, and $\beta$ is the
average batch size and is specified by the user.

The parameters for the procedure are as follows. Before the
procedure initiates, the user selects an overall confidence level
$1-\alpha$, type-I error rates $\alpha_1$, $\alpha_2$ such that
$\alpha=\alpha_1+\alpha_2$, an indifference-zone parameter $\delta$,
Stage~0 and Stage~1 sample sizes $n_0,n_1\ge 2$, and average Stage~2
batch size $\beta$. The user also chooses $\bar{r}>0$ as the maximum
number of iterations in Stage~2, which governs how much simulation
budget to spend in iterative screening before moving to
indifference-zone selection in Stage~3.

Typical choices for error rates are $\alpha_1 = \alpha_2 = 0.025$ for guaranteed PGS
of 95\%. The indifference-zone parameter $\delta$ is usually chosen within the context
of the application, and is often referred to as the smallest
difference worth detecting. The sample sizes $n_0$ and $n_1$ are typically chosen to be small multiples of 10, with the view that these
give at least reasonable estimates of the runtime per replication and
the variance per replication.

For non-normal simulation output, we recommend setting $\beta\ge 30$
to ensure normally distributed batch means. The parameter $\beta$ also
helps to control communication frequency so as not to overwhelm
the master with messages. Let $T_\text{sim}$ be a crude estimate of
the average simulation time (in seconds) per replication, perhaps
obtained in a debugging phase. Then ideally
the master communicates with a worker every $\beta T_\text{sim}/c$
seconds. If every communication takes $T_\text{comm}$ seconds,
the fraction of time the master is busy is
$\rho=cT_\text{comm}/\beta T_\text{sim}$. We recommend setting $\beta$ such
that $\rho\le 0.05$, in order to avoid significant waiting of workers.

We recommend choosing $\bar{r}$ such that a fair amount of simulation
budget (no more than 20\% of the sum of Rinott sample sizes) will be
spent in the iterative screening stages. Note that a small $\bar{r}$
implies insufficient screening whereas a large $\bar{r}$ may be too
conservative.

Under these general principles, our choices of
$(\beta=100,\bar{r}=10)$ and $(\beta=200,\bar{r}=5)$ in the
experiments in \S\ref{sect:computation} work reasonably
well on our testing platform, but it is conceivable that other values
could improve performance.

Finally, we define some quantities used in the iterative screening stages. Let $\eta$ be the solution to
%\begin{align}% Old version 
%E\left[2\sum_{s=0}^{\infty}(-1)^s\bar{\Phi}\left(\frac{a(2s+1)}{\sqrt{t(n_1-1)}}\sqrt{R}\right)\right]=1-(1-\alpha_1)^{\frac{1}{k-1}}. \label{eq:Choice_eta}
%\end{align}
\begin{align}% New version
E\left[2\bar{\Phi}\left(\eta\sqrt{R}\right)\right]=1-(1-\alpha_1)^{\frac{1}{k-1}}, \label{eq:Choice_eta}
\end{align}
where $\bar{\Phi}(\cdot) $ denotes the complementary standard normal distribution function, and $R$ is the minimum of two i.i.d. $\chi^2$ random variables, each with $n_1-1$ degrees of freedom.  Let the distribution function and density of such a $\chi^2$ random variable be denoted $F_{\chi^2_{n_1-1}}(x)$ and $f_{\chi^2_{n_1-1}}(x)$, respectively. Hence $R$ has density
$
f_R(x)=2[1-F_{\chi^2_{n_1-1}}(x)]f_{\chi^2_{n_1-1}}(x).
$
Also, for any two systems $i\ne j$, define 
\begin{align*}
t_{ij}(r)&=\left[\frac{\sigma^2_i}{n_i(r)}+\frac{\sigma^2_j}{n_{j}(r)}\right]^{-1},
&&Z_{ij}(r)=t_{ij}(r) [\Xbar_i(n_i(r))-\Xbar_{j}(n_{j}(r))], 
\\
\tau_{ij}(r)&=\left[\frac{S^2_i}{n_i(r)}+\frac{S^2_{j}}{n_{j}(r)}\right]^{-1},
&&Y_{ij}(r)=\tau_{ij}(r) [\Xbar_i(n_i(r))-\Xbar_{j}(n_{j}(r))],
\\
a_{ij}(\bar{r}) &= \eta \sqrt{(n_1-1)\tau_{ij}(\bar{r})}.
\end{align*}
%We will show later that $Z_{ij}(r)$ and $Y_{ij}(r)$ are related to a Brownian motion with drift $\mu_i-\mu_j$, $t_{ij}(r)$ and $\tau_{ij}(r)$ behave like ``time units'' of the Brownian motion scaled by true and estimated system variances, respectively, and $a_{ij}(\bar{r})$ and $\tau_{ij}(\bar{r})$ jointly form a rectangular continuation region used in Stages~1 and 2. If the Brownian motion exists the continuation before reaching $\tau_{ij}(\bar{r})$ then one of systems $i$ and $j$ is eliminated, otherwise, we stop screening and switch to indifference-zone selection in Stage~3.

\subsection{Good Selection Procedure under Unknown Variances} \label{sect:GSPsteps}
%The procedure is briefly stated as follows. Interested readers may refer to the repository \citep{MPIRNS} for its C++ implementation and documentation.
%The required inputs to the procedure are type-I error rates $\alpha_1$, $\alpha_2$, sample size used for estimating completion time, $n_0$, and for estimating system variance, $n_1$, average batch size $\beta $, maximum number of screening batches $\bar{r}$
\begin{enumerate}
    
    \item \textbf{(Stage~0), optional} Master assigns systems to workers, so that each system $i$ is simulated for $n_0$ replications and the average simulation completion time $\bar{T_i}$ is reported to the master.
    
    \item \textbf{(Stage~1)} Master assigns systems to load-balanced (using $\bar{T_i}$ if available) simulation groups $G_1^w$ for $w = 1,\ldots, c$. Let $\mcI\leftarrow\mcS$ be the set of surviving systems.
    
    \item \label{step:Stage1Main} For $w=1,2,\ldots,c$ in parallel on workers: 
    
    \begin{enumerate}[label=(\alph{enumii})]
        \item Sample $X_{i\ell}$, $\ell=1,2,\ldots, n_1$ for all $i\in G_1^w$.
        \item Compute Stage~1 sample means and variances $\Xbar_i(n_1)$ and $S^2_i$ for $i\in G_1^w$. 
        \item \label{step:NHHStage1Screening} Screen within group $G_1^w$: system~$i$ is eliminated (and removed from $\mcI$) if there exists a system~$j\in G_1^w: j\neq i$ such that 
        %\eric{$\tau_{ij}(0)<\tau_{ij}(\bar{r})$ and }
        $
        Y_{ij}(0)<
        %\min[0,
        -a_{ij}(\bar{r})
        %]
        $.
        \item Report survivors, together with their Stage~1 sample means $\Xbar_i(n_i(0))$ and variances $S_i^2$, to the master.
    \end{enumerate}
    
    \item \textbf{(Stage~2)} Let $G_1\leftarrow\mcI$ be the set of systems surviving Stage~1.
    Master computes sampling rule \eqref{eq:sampling_rule} using $S_i^2$ obtained in Stage~1, and divides $G_1$ to approximately load-balanced screening groups $G^w_2$ for $w = 1,\ldots, c$. Let $s_i\leftarrow 0, i\in G_1$ be the count of the number of batches simulated in Stage~2 for system~$i$.
    
    \item \label{step:Stage2Main} For $w=1,2,\ldots,c$ in parallel on workers, let $r_w \leftarrow 0$ be the count of the number of batches screened on worker~$w$ (which is common to all systems in the screening), and iteratively switch between simulation and screening as follows
    (this step entails some communication with the master, the details of which are omitted): 
    
    \begin{enumerate} [label=(\alph{enumii})]
        \item \label{step:CheckS1}
        Check termination criteria with the master: 
        if $|\mcI|=1$ (only one system remains) or $r_w\ge\bar{r}$ for all $w$ (each worker has screened up to $\bar{r}$, the maximum number of batches allowed), go to Stage~3; otherwise continue to Step~\ref{step:Stage2Main}\ref{step:checkSim}. 
        %{seems like $\mcS$ is overloaded to denote both the initial set of systems and the remaining set of systems?}\susan{I agree. Eric, please change this.}
        \item \label{step:checkSim}
        Decide to either simulate more replications or perform screening based on available results: 
        check with the master if the $(r_w+1)$th iteration has completed for all systems $i\in G^w_2$ and $|G^w_2|>1$, if so, go to Step~\ref{step:Stage2Main}\ref{step:NHHscreening}, otherwise go to Step~\ref{step:Stage2Main}\ref{step:NHHsim}.
        \item \label{step:NHHsim} 
        Retrieve the next system~$i$ in $G_1$ (not necessarily $G^w_2$) from the master and simulate it for an
        additional $n_i(s_i+1)-n_i(s_i)$ replications. Set $s_i\leftarrow s_i+1$. Report simulation results to the master. Go to Step~\ref{step:Stage2Main}\ref{step:CheckS1}.
        \item \label{step:NHHscreening} 
        Screen within  $G^w_2$ as follows. Retrieve necessary statistics for systems in $G^w_2$ from the master (recall that a system in $G^w_2$ is not necessarily simulated by worker $w$). Let $r_w
        \leftarrow r_w+1$. A system~$i\in G^w_2$ is eliminated if $r_w\le
        \bar{r}$ and there exists a system~$j\in G^w_2:j\neq i$ such that 
        %\eric{$\tau_{ij}(r_w)<t$ and }
        $
        Y_{ij}(r_w)
        <
        %\min[0,
        -a_{ij}(\bar{r})
        %+\lambda \tau_{ij}(r_w) 
        %]
        $. 
        %where
        %$
        %\tau_{ij}(r_w)
        %=[S_i^2/n_i(r_w)+S_j^2/n_j(r_w)]^{-1}$.
        %\item \label{step:NHHscreenother} 
        Also use a subset of systems from other workers, e.g., those with the highest sample mean from each worker, to eliminate systems in $G^w_2$.
        %\item \label{step:NHHfinal} 
        Remove any eliminated system from $G^w_2$ and $\mcI$. Go to Step~\ref{step:Stage2Main}\ref{step:CheckS1}.
    \end{enumerate}
    
    \item \textbf{(Stage~3)} 
    Let $G_2\leftarrow\mcI$ be the set of systems surviving Stage~2.
    %Let $G_2:=\{\text{System }i: i\text{ survives Stages~1~and~2} \}$. 
    If $k':=|G_2|=1$, select the single system in $G_2$ as the best. Otherwise, set 
    %$t=t_{(1-\alpha_0)^{\frac{1}{k-1}}, n_1-1}$ and
    $h=h(1-\alpha_2,n_1,k')$, where the function $h(\cdot)$ gives Rinott's constant (see e.g. \citealp[Chapter 2.8]{Bechhofer1995}). For each remaining system $i\in G_2$, compute
    $
    N_i=\max\{ n_i(\bar{r}),\lceil (hS_i/\delta)^2 \rceil \}, 
    %\label{eq:Rinott}
    $
    and take an additional $\max\{N_i-n_i(\bar{r}),0\}$ sample observations in parallel.
    Once a total of $N_i $ replications have been collected in Stages~1 through 3 for each $i\in G_2$, select the system $K$ with the highest $\Xbar(N_K)$ as the best.
    %\item Report the single surviving system in $\mcS$ as the best.
\end{enumerate}

\subsection{Guaranteeing Good Selection}

Our probabilistic guarantee on the final solution relies on the following assumption on the distribution of simulation output, which is common to the sequential R\&S literature.
\begin{assumption}\label{assum:mvn}
    For each system $i = 1,2,\ldots,k$, the simulation output random variables 
    $\{X_{i\ell}, \ell = 1,2,\ldots\}$ 
    are i.i.d. replicates of a random variable $X_i$ having a normal distribution with finite mean $\mu_i$ and finite variance $\sigma^2_i$, and are mutually independent for different $i$.
\end{assumption}
In \S\ref{sect:RandCompTime} we gave conditions under which the simulation output generated by parallel cores satisfies the i.i.d. assumption. 

We now formally state our good selection guarantee.
\begin{theorem} \label{theorem:GoodSelection}
    Under Assumption~\ref{assum:mvn}, GSP selects a system $K$ that satisfies $\mu_k-\mu_K\le\delta$
    with probability at least $1-\alpha$.
\end{theorem}

We discuss the key insights that yield the PGS guarantee here and
defer the full proof of Theorem~\ref{theorem:GoodSelection} to
\S\ref{sect:GSPproof}.

\begin{proof}[Sketch of Proof]
    
%\proof{Sketch of Proof}
First, we show that the best system, System $k$, survives the
iterative screening in Stages~1 and 2 with probability at least
$1-\alpha_1$, irrespective of whether the best solution is unique or
not. Indeed, conditioning on the Stage~1 variance estimates
$\{S_i^2:1=1,2,\ldots,k \}$, we can, for any system $i\ne k$, relate
the batch statistics $Z_{ki}(r):r=0,1,\ldots, \bar{r}$ to a properly
scaled Brownian motion with drift $\mu_k-\mu_i\ge 0$. Then, using the
reflection principle of Brownian motion, we can upper-bound the
probability that the scaled Brownian motion falls below some number
$-a$ before some time $t$, or equivalently, the probability that
$Y_{ki}(r)$ falls below $-a_{ki}(\bar{r})$ in some $r$th iteration
where $r\le\bar{r}$, which is the criterion used to eliminate system
$k$ in the iterative screening stages.
The construction of continuation region parameter $\eta$ and the fact that $(n_1-1)S_i^2/\sigma_i^2\sim\chi^2_{n_1-1}$ for all $i$ jointly ensure that the unconditional probability of eliminating $k$ is no greater than $\alpha_1$.

Second, as Stage~3 is closely related to the \cite{Rinott1978}
procedure with confidence level $1-\alpha_2$, it follows from
Theorem~1 of \cite{NM1995} that 
$$P\left\lbrace\bar{X}_K(N_K) -
\bar{X}_i(N_i) - \left( \mu_K-\mu_i \right)>-\delta, \forall i\in
G_2: i\ne K \right\rbrace\ge 1-\alpha_2$$ where $K$ is the system
with the highest sample mean after Stage~3. Therefore we conclude that
if $k\in G_2$  (the best system~$k$ survives Stages~1 and 2), then
$P\left\lbrace\bar{X}_K(N_K) - \bar{X}_k(N_k) - \left( \mu_K-\mu_k
\right)>-\delta \right\rbrace\ge 1-\alpha_2$, that is, Stage~3
selects a good system $K$ such that $\mu_K\ge \mu_k-\delta$ with high
probability. 

Finally, we complete the proof by invoking a result from
\cite{Nelson2001} that guarantees an overall PGS of
$1-\alpha_1-\alpha_2$ for two-stage procedures. 

\end{proof}
%\endproof

The key difference between the screening methods used in GSP and the
$\mathcal{KN}$ family procedures \citep{KimNels01,HN2005,Hong2006} is
that  the $\mathcal{KN}$ family relies on the PCS assumption
($\mu_k-\mu_i\ge \delta > 0$ for all $i\ne k$) to guarantee PCS,
whereas our approach does not. Therefore, GSP works for any
indifference-zone parameter $\delta>0$, and when there exist multiple
systems $i$ such that $\mu_i\ge\mu_k-\delta$, GSP is guaranteed to
select one such system with probability at least $1-\alpha$.

\subsection{Choice of parameter $\eta$} \label{sect:compute_eta}
%Similar to the choice of sampling rules, $a$ and $t$ are allowed to depend on Stage~1 variances $S^2_i$ without altering the normality of $X$.
%
%Choosing $t$:
%\begin{itemize}
%\item The actual sample sizes from systems~$i$~and~$j$ for $Z_{ij}(t_{ij}(r))$ to reach $t$ depend only on $S^2_i$, $S^2_j$ and $t$. After Stage~1 we may estimate this sample size as well as the Rinott sample size $N_i$ given in \eqref{eq:Rinott}. 
%\item We may set $t$ such that the maximum sample size used in Stage~2 is a fraction of the Rinott sample size $N_i$.
%\item Alternatively, we may set $t$ such that by the end of Stage~2, at least $N_i$ samples have been collected for each surviving system so the Stage~3 sampling can be eliminated.
%\end{itemize}
%Choosing $a$ given $t$:
%\begin{itemize}
%\item Equation~\eqref{eq:Choice_eta} is upper-bounded by
%\begin{align}
%E\left[ 2\bar{\Phi}\left(\frac{a}{\sqrt{t(n_1-1)}}\sqrt{R} \right)\right] \label{eq:Choice_a_approx}
%\end{align}
%and for any given $t$, setting \eqref{eq:Choice_a_approx} to $1-(1-\alpha_1)^{\frac{1}{k-1}}$ and solving for $a$, the solution $a_u$ is no less than the solution $a$ to \eqref{eq:Choice_eta}. Using $a_u$ instead of $a$ is more conservative and does not reduce the PGS.
%
%Let $\eta=\frac{a}{\sqrt{t(n_1-1)}}$, it follows that
One way to compute $\eta$, the solution to \eqref{eq:Choice_eta}, is by integrating the LHS using Gauss-Laguerre quadrature and using bisection to find the root of \eqref{eq:Choice_eta}. Alternatively, we may employ a bounding technique to approximate $\eta$ as follows. Indeed, the LHS of \eqref{eq:Choice_eta} is
\begin{align}
E\left[2\bar{\Phi}\left(\eta\sqrt{R}\right)\right]
&=\int_{y=0}^{\infty}2\bar{\Phi}(\eta\sqrt{y}) 2 [1-F_{\chi^2_{n_1-1}}(y)] f_{\chi^2_{n_1-1}}(y)\mathrm{d}y \nonumber
\\&\le \int_{y=0}^{\infty}4\bar{\Phi}(\eta\sqrt{y}) f_{\chi^2_{n_1-1}}(y)\mathrm{d}y \label{eq:CDFUpperBound}
\\&\le
%\int_{y=0}^{\infty}4\frac{e^{-\eta^2y/2}}{\eta\sqrt{2\pi y}} f_{\chi^2_{n_1-1}}(y)\mathrm{d}y
%\\&=
\int_{y=0}^{\infty}4\frac{e^{-\eta^2y/2}}{\eta\sqrt{2\pi y}} \frac{y^{\frac{n_1-1}{2}-1} e^{-y/2}}{ 2^{\frac{n_1-1}{2}} \Gamma(\frac{n_1-1}{2})}\mathrm{d}y \label{eq:PhiBarApprox}
\\&= \frac{4\Gamma(\frac{n_1-2}{2})(\frac{2}{\eta^2+1})^{\frac{n_1-2}{2}}}{\sqrt{2\pi}\eta 2^{\frac{n_1-1}{2}}\Gamma(\frac{n_1-1}{2})}
\int_{0}^{\infty} \frac{(\frac{\eta^2+1}{2})^{\frac{n_1-2}{2}} y^{\frac{n_1-1}{2}-1-\frac{1}{2}} e^{-\frac{\eta^2+1}{2}y}}{\Gamma({\frac{n_1-2}{2}})} \mathrm{d}y\label{eq:GammaPDF}
\\&= \frac{2\Gamma(\frac{n_1-2}{2})}{\sqrt{\pi}\Gamma(\frac{n_1-1}{2})\eta ({\eta^2+1})^{\frac{n_1-2}{2}} }, \label{eq:bound_eta}
\end{align}
where \eqref{eq:CDFUpperBound} holds because distribution functions are non-negative and is inspired by a similar argument in \cite{Hong2006}, \eqref{eq:PhiBarApprox} follows from the fact that $\bar{\Phi}(x)\le {e^{-x^2/2}}/({x\sqrt{2\pi}})$ for all $x>0$, and the integrand in \eqref{eq:GammaPDF} is the pdf of a Gamma distribution with shape $(n_1-1)/2$ and scale $2/(\eta^2+1)$, and hence integrates to 1.

Note that \eqref{eq:bound_eta} is an upper-bound on the left-hand side of \eqref{eq:Choice_eta}. Setting \eqref{eq:bound_eta} to $1-(1-\alpha_1)^{\frac{1}{k-1}}$ and solving for $\eta$ yields an overestimate $\eta'$, which is more conservative and does not reduce the PGS. Furthermore, as \eqref{eq:bound_eta} is strictly decreasing in $\eta$, $\eta'$ can be easily determined using bisection.

The parameter $\eta$ determines the value of $a_{ij}(\bar{r})$, and hence how quickly an inferior system is eliminated in screening Steps~\ref{step:Stage1Main}\ref{step:NHHStage1Screening} and \ref{step:Stage2Main}\ref{step:NHHscreening}. The value of $\eta$ therefore directly impacts the effectiveness of the iterative screening. Hence, it is desirable that $\eta$ does not grow dramatically as the problem gets bigger. Observe that \eqref{eq:bound_eta} can be further bounded by
\begin{align}
%\eqref{eq:bound_eta}
%\le 
\frac{2\Gamma(\frac{n_1-2}{2})}{\sqrt{\pi}\Gamma(\frac{n_1-1}{2})\eta ^{n_1-1} } :=C\eta^{1-n_1}. \label{eq:crude_eta}
\end{align}
Setting \eqref{eq:crude_eta} to $1-(1-\alpha_1)^{\frac{1}{k-1}}$ implies that the right-hand side of \eqref{eq:crude_eta} must be small. After some further manipulations we have 
\begin{align}
\log(1-\alpha_1 )=(k-1)\log \left( 1-C\eta^{1-n_1} \right) \approx (k-1)(-C\eta^{1-n_1}) 
\label{eq:k_eta_relation}
\end{align}
where the approximation holds because $\log(1-\epsilon)\approx -\epsilon$ for small $\epsilon>0$. It follows from \eqref{eq:k_eta_relation} that for fixed $\alpha_1$, 
% and sufficiently large values of $n_1$ and $k$, 
the parameter $\eta$ grows very slowly with respect to $k$, at a rate of $k^{1/(n_1-1)}$. Therefore, the continuation region defined by $\eta$ and $\bar{r}$ as well as the power of our iterative screening are not substantially weakened as the number of systems increases, especially when $n_1$ or $k$ is large. 
In this regime, we should expect the total cost of this R\&S procedure to grow approximately linearly with respect to the number of systems.

%\item 
%A Brownian motion with drift $\Delta$ exits $[-a,a]$ via the upper boundary (reaching $a$ before $-a$) before time $t$ with probability \citep{Hall1997}
%\begin{align}
%P_\Delta(t,U)=\sum_{j=0}^{\infty}(-1)^je^{a\Delta}[e^{a(2j+1) \Delta} \bar{\Phi}(\frac{a(2j+1)+\Delta t}{\sqrt{t}})+e^{-a(2j+1) \Delta} \bar{\Phi}(\frac{a(2j+1)-\Delta t}{\sqrt{t}})] \label{eq:BMwithDrift}
%\end{align}
%Assuming our objective is to eliminate systems outside $\Delta$ of the best, an intuitively sound choice of $(a,t)$ maximizes $P_\Delta(t,U)$ subject to \eqref{eq:Choice_eta}. As an approximation, let $\gamma=a/\sqrt{t}$,
%
%\begin{align}
%\text{The first term in \eqref{eq:BMwithDrift}}
%&=e^{a\Delta}
%[e^{a\Delta}\bar{\Phi}(\gamma+a\Delta/\gamma)+
%e^{-a\Delta}\bar{\Phi}(\gamma-a\Delta/\gamma)] 
%\\&\lesssim e^{2a\Delta} 
%\phi(\gamma+a\Delta/\gamma)+
%\phi(\gamma-a\Delta/\gamma)
%=\frac{\sqrt{2}}{\sqrt{\pi}} e^{-(\gamma-a\Delta/\gamma)^2/2}
%\end{align}
%which is maximized when $(\gamma-a\Delta/\gamma)^2 $ is minimized at $a^*=t\Delta$. \eric{TBD: Offer intuition here.}
%%\end{itemize}

\section{Computational Study}%================================================== 
\label{sect:computation}

In this section, we discuss our parallel computing environment and test problem, discuss two parallel implementations of GSP, and discuss the results of our numerical experiments. 

\subsection{Parallel Computing Environment and Test Problem}
Our numerical experiments are conducted on Extreme Science and Engineering Discovery Environment (XSEDE)'s Stampede high-performance cluster. The Stampede cluster contains over 6,400 computer nodes, each equipped with two 8-core Intel Xeon E5 processors and 32 GB of memory and runs a Linux Centos 6.3 operating system \citep{TACCStampede}. Parallel programs are submitted through the Simple Linux Utility for Resource Management (SLURM) batch environment which enables users to specify the number of cores to use. The high-performance processors on Stampede are highly reliable and we have never seen a core failure.

We test R\&S procedures on a throughput-maximization problem taken from
\texttt{SimOpt.org} \\\citep{simoptlib}. In this problem, we solve 
\begin{align}
\max_x &\,\, E[g(x;\xi)] \label{eq:TpMaxObj}
\\\text{s.t. } r_1+r_2+r_3&=R \nonumber
\\ b_2+b_3&=B \nonumber
\\ x = (r_1,r_2,r_3,b_2,b_3) &\in \{1, 2, \ldots\}^5 \nonumber
\end{align}
where the function $g(x;\xi)$ represents the random throughput of a three-station flow line with finite buffer storage in front of Stations
2 and 3, denoted by $b_2$ and $b_3$ respectively, and an infinite number of jobs in front of Station 1. The processing times of each job at stations 1, 2, and 3 are independently exponentially distributed with service rates $r_1$, $r_2$ and $r_3$, respectively. The overall objective is to maximize expected steady-state throughput by finding an optimal (integer-valued) allocation of buffer and service rate.

For each choice of the problem parameters $R,B\in \mathbb{Z}^+$, the
number of feasible allocations is finite and can be easily computed. We consider three problem instances with very different sizes presented in Table~\ref{tab:TpMax}. Since the service times are all exponential, we can analytically compute the expected throughput of each feasible allocation by modeling the system as a continuous-time Markov chain. Furthermore, it can be shown that $E[g(r_1,r_2,r_3,b_2,b_3;\xi)] = E[g(r_3,r_2,r_1,b_3,b_2;\xi)] $ for any feasible allocation $(r_1,r_2,r_3,b_2,b_3)$, so the problem may have multiple optimal solutions. Therefore, this is a problem for which the PCS assumption $\mu_k-\mu_{k-1}\ge \delta>0$ can be violated and R\&S procedures that only guarantee correct selection might be viewed as heuristics. 

\begin{table}
    \centering
    \caption{Summary of three instances of the throughput maximization problem.}
    \renewcommand{\arraystretch}{0.5}
    \begin{tabular}
        %{0.99\textwidth}
        {rr r *{3}{C{0.07\textwidth}} *{2}{C{0.085\textwidth}} *{1}{C{0.07\textwidth}}}
        %\hline
        \toprule
        & Number of & Highest & \multicolumn{3}{c}{$p$th percentile of system means} & \multicolumn{3}{c}{No. of systems in $[\mu_k-\delta,\mu_k]$} \\
        %\cmidrule(lr){4-6} \cmidrule(l){7-9} 
        $(R,B)$ & systems $k$ & mean $\mu_k$ & $p=75$ & $p=50$ & $p=25$ & $\delta=0.01$ & $\delta=0.1$ & $\delta=1$ \\
        \midrule 
        $(20,20)$ & 3,249 & 5.78 & 3.52 & 2.00 & 1.00 & 6 & 21 & 256 \\
        $(50,50)$ & 57,624 & 15.70 & 8.47 & 5.00 & 3.00 & 12 & 43 & 552 \\
        $(128,128)$ & 1,016,127 & 41.66 & 21.9 & 13.2 & 6.15 & 28 & 97 & 866 \\
        \bottomrule
    \end{tabular}%
    \renewcommand{\arraystretch}{1}
    \label{tab:TpMax}%
\end{table}%

%Table~\ref{tab:TpMax} shows that, in each of the three problem instances, there are many near-optimal systems as well as many systems that are clearly sub-optimal. The systems also differ significantly in variances \citep{Ni2014}.

By default, in each simulation replication, we warm up the system for
2,000 released jobs starting from an empty system, before observing
the simulated throughput to release the next 50 jobs. This may not be
the most efficient way to estimate steady-state throughput compared to
taking batch means from a single long run, but it suits our purpose
which is to obtain i.i.d. random replicates from the $g(x;\xi)$
distribution in parallel. Due to the fixed number of jobs, the
wall-clock time for each simulation replication exhibits low
variability.

\subsection{Parallel Implementations of GSP} \label{sect:implementation_parallel}
In this section, we discuss two implementations of GSP proposed in \S\ref{sect:GoodSelectionProcedure}. Although we will primarily test them on Stampede, both procedures can be configured to run on a wide range of parallel platforms from multi-core personal computers to the Amazon EC2 cloud.

\subsubsection{MPI} \label{sect:implementation_mpi} 
{
    Message-Passing Interface (MPI) is a popular distributed-memory parallel programming framework with libraries available in C/C++ and Fortran and the de-facto standard for parallel programming on many high-performance parallel clusters including Stampede. Using MPI, programs operate in an environment where Assumptions 1 and 2 hold. The method by which parallel cores independently execute instructions and communicate through message-passing can be highly customized to serve different purposes. 
}

The MPI implementation is a realization of GSP presented in
\S\ref{sect:GoodSelectionProcedure}, and is fully described in
\S\ref{sect:FullImplementation_MPI}. We designate one core as
the master and let it control other worker cores. We observe that
communication is fast on Stampede, taking only $10^{-6}$ to $10^{-4}$
seconds each time depending on the size of the message. Therefore,
with an appropriate choice of the batch-size parameter $\beta$, the
master remains idle most of the time so the workers are usually able
to communicate with the master without much delay.

Our MPI implementation is designed primarily for high-performance
clusters like Stampede and does not detect and manage core
failures. As simulation output is distributed across parallel cores
without backup, the MPI implementation is vulnerable to core failures
which may cause loss of data and break the program. {In practice, such failures can occur for a number of reasons, including faulty hardware but also cores aborting due to being re-assigned to higher priority tasks by the system.} Therefore, for
cheap and less reliable parallel platforms, the MPI implementation
needs to be augmented with a ``fault-tolerant'' mechanism in order to
allow the procedure to continue even if some cores fail. This motivates us to seek alternative programming tools such as MapReduce that handle core failures automatically.
{Our MPI implementation uses the mvapich2 library and its source code and documentation is hosted
    in the open-access repository \cite{MPIRNS}. }

%highly customized for the application, Explicit message-passing, all operations in-memory

\subsubsection{Hadoop MapReduce} \label{sect:implementation_hadoop}
MapReduce \citep{Dean2008} is a distributed computing model typically
used to process large amounts of data. Conceptually, each MapReduce
instance consists of a \textit{Map} phase where \textit{data entries}
are processed by ``Mapper'' functions in parallel, and a
\textit{Reduce} phase where Mapper outputs are grouped by
\textit{keys} and summarized using parallel ``Reducer'' functions. {
    MapReduce has become a popular choice for data intensive applications such as PageRank and TeraSort, thanks to the following advantages.
    \begin{itemize}
        \item \textbf{Simplicity. }The MapReduce programming model allows its users to solely focus on designing meaningful Mappers and Reducers that define the parallel program, without explicitly handling the complex details of the message-passing and the distribution of workload to cores, a task which is completely delegated to the MapReduce package.
        \item \textbf{Portability. }Hadoop is a highly popular and portable MapReduce system that can be easily deployed with minimal changes on a wide range of parallel computer platforms such as the Amazon EC2 cloud. 
        \item \textbf{Resilience to core failures. }On less reliable hardware where there is a non-negligible probability of core failure, the Apache Hadoop system can automatically reload any failed Mapper or Reducer job on another worker to guide the parallel job towards completion. 
    \end{itemize}
    
    Despite these advantages, the use of MapReduce for computationally intensive and highly iterative applications, such as simulation and simulation optimization, is less documented. Moreover, most popular MapReduce implementations such as Apache Hadoop have limitations that may potentially reduce the efficiency of highly iterative algorithms such as parallel R\&S procedures.
    \begin{itemize}
        \item \textbf{Synchronization. }By design, each Reduce phase cannot start before the previous Map phase finishes and each new MapReduce iteration cannot be initiated unless the previous one shuts down completely. Hence, a R\&S procedure using MapReduce has to be made fully synchronous. If load-balancing is difficult, for instance as a result of random simulation completion times, then core hours could be wasted due to frequent synchronization. 
        \item \textbf{Absence of Cache. } 
        In Apache Hadoop, workers are not allowed to cache any information between Map and Reduce phases. 
        As a result, the outputs generated by Mappers and Reducers are often written to a distributed file system (which are usually located on hard drives) before they are read in the next iteration. Compared to the MPI implementation where all intermediate variables are stored in memory, the MapReduce version could be slowed down by repeated data I/O.
        Moreover, the lack of cache requires the simulation program, including any static data and/or random number generators, to be initialized on workers before every MapReduce iteration.
        \item \textbf{Nonidentical Run Times. }
        By default, Apache Hadoop does not dedicate each worker to a single task. It may simultaneously launch several Mappers and Reducers on a single worker, run multiple MapReduce jobs that share workers on the same cluster, or even use workers that have different hardware configurations. 
        In any of theses cases, simulation completion times may be highly variable and time-varying. 
        Therefore, Stage~0 of GSP (estimation of simulation run time) is dropped from our MapReduce implementation.
    \end{itemize}
}

\begin{table}
    \small\centering
    \caption{Major differences between GSP implementations}
    \renewcommand{\arraystretch}{1}
    \begin{tabular}{|>{\raggedright\arraybackslash}p{0.25\textwidth}|>{\raggedright\arraybackslash}p{0.35\textwidth}|>{\raggedright\arraybackslash}p{0.35\textwidth}|}
        \hline%\toprule
        Task & MPI & Hadoop MapReduce \\
        \hline \hline
        Master & Explicitly coded & Automated \\\hline
        Message-passing & Explicitly coded & Automated \\\hline
        Synchronization & Once after each stage & More frequent: required in every \\
        &&iteration of Stage~2 \\\hline
        Simulation & Each worker simulates one system  & Each worker simulates multiple  \\
        &per iteration &systems per iteration\\\hline
        Load-balancing & Via asynchronous & By assigning approximately equal  \\
        &communications between the & number of systems (Mappers) to each \\
        &master and a single worker  & worker in each synchronized iteration\\\hline
        Batch statistics and  & Always stored in workers' memory & Written to hard disk after each  \\
        random number seeds &&iteration\\\hline
        %\bottomrule
    \end{tabular}%
    \renewcommand{\arraystretch}{1}
    \label{tab:MPIvHadoop_Implementations}%
\end{table}%

Although there are specialized MapReduce variants such as ``stateful
MapReduce'' that attempt to address these limitations
\citep{Elgohary2012}, we do not explore them as they are less
available for general parallel platforms, at least at present. {However, some of these limitations (such as the lack of caching across multiple MapReduce rounds) are idiosyncratic to specific packages like Hadoop rather than the framework itself. Nevertheless, the a priori expectation is that, for a highly iterative procedure like ours, a highly optimized MPI approach will outperform a Hadoop one; thus our question is not which is fastest, but whether MapReduce can offer most of the speed of MPI along with its advantages discussed above.}

%Traditionally, MapReduce programs efficient on data intensive applications such as PageRank and TeraSort. 
%However, its use on computationally intensive applications, such as simulation and simulation optimization, is less documented. 

%We investigate in this section the easiness and efficiency of solving R\&S problems using MapReduce. 

We propose a variant of GSP using iterative MapReduce as follows. In
each Mapper function, we treat each surviving system as a single data
entry, obtain an additional batched sample, and output updated summary
statistics such as sample sizes, means, and variances. Each output
entry is associated with a key which represents the screening group to
which it belongs. Once output entries of Mappers are grouped by their
keys, each Reducer receives a group of systems, screens amongst them,
and writes each surviving system as a new data entry which in turn is
used as the input to the next Mapper.

To fully implement GSP, MapReduce is run for several iterations. The
first iteration implements Stage~1, where both $\Xbar_i(n_1)$ and
$S^2_i$ are collected. Then, a maximum number of $\bar{r}$ subsequent
iterations are needed for Stage~2, with only $n_i(r)$ and
$\Xbar_i(n_i(r))$ being updated in each iteration. (Additional
MapReduce iterations can be run where the best system from each group
is shared for additional between-group screening.) The same Reducer
can be applied in both Stages~1 and 2, as the screening logic is the
same. Finally, a Stage~3 MapReduce features a Mapper that calculates
the additional Rinott sample size, simulates the required
replications, and a different Reducer that simply selects the system
with the highest sample mean at the end. Full details of each stage
are provided in \S\ref{sect:FullImplementation_Hadoop}.

Our MapReduce implementation is based on the native Java interface for
MapReduce provided in Apache Hadoop 1.2.1. It is hosted
in the open-access repository \cite{MAPREDRNS}.  Table~\ref{tab:MPIvHadoop_Implementations} summarizes some of the major differences between the MPI and Hadoop implementations. 
%{to save space, we could move table 2 to the appendix}

\subsection{Numerical Experiments}
We now demonstrate the practical performance of GSP by using it to solve the throughput maximization test problem.

\subsubsection{GSP vs Existing Parallel Procedures} \label{sect:GSPvNHHvNSGS}

\begin{table}
    \caption{A comparison of procedure costs using parameters $n_0=20$,
        $n_1=50$, $\alpha_1=\alpha_2=2.5\%$, $\beta=100$, $\bar{r} =
        10$. (Results to 2
        significant figures)}
    \centering
    %\linespread{1}
    %\onehalfspacing
    \renewcommand{\arraystretch}{1}
    \begin{tabular}
        %{\textwidth}
        {lllrr}
        \toprule
        Configuration & $\delta$ & Procedure & Wall-clock time (sec) & Total number of \\
        & & &  &  simulation replications ($\times 10^6$)\\
        %& & & time (s) &  \\
        \midrule 
        3,249 systems  & 0.01  & GSP   &                       14  &                                         2.3  \\
        on 64 cores &       & NHH   &                       14  &                                         2.5  \\
        &       & \NSGS\  &                    120  &                                      13  \\ \cline{2-5}
        & 0.1   & GSP   &                         3.4  &                                            0.57  \\ 
        &       & NHH   &                         2.6  &                                            0.44  \\
        &       & \NSGS\  &                         3.4  &                                            0.48  \\
        \hline
        57,624 systems & 0.01  & GSP   &                    720  &                                    130  \\
        on 64 cores &       & NHH   &                    520  &                                      89  \\
        &       & \NSGS\  &              11,000  &  1600  \\\cline{2-5}
        & 0.1   & GSP   &                       60  &                                      10  \\
        &       & NHH   &                       71  &                                      12  \\
        &       & \NSGS\  &                    150  &                                      23  \\
        \hline
        1,016,127 systems & 0.1   & GSP   &                    260  &                                    320  \\
        on 1,024 cores &       & NHH   &                 1,000  &                                    430  \\
        &       & \NSGS\  &                 1,400  & 1900 \\
        \bottomrule
    \end{tabular}%
    \renewcommand{\arraystretch}{1}
    \label{tab:GSPvNHHvNSGS}%
\end{table}%

GSP is motivated by an earlier computational study by \cite{Ni2014},
which compares the performance of two parallel procedures, NHH
\citep{Ni2013} and \NSGS\ \citep{Ni2014}. NHH is a parallel procedure
that adopts the fully-sequential serial procedure proposed in
\cite{Hong2006}, and only provides a PCS guarantee \citep{HN2014}. For
problems where multiple optimal solutions exist, it is used as a
heuristic because the PCS assumption does not hold. NHH can 
be viewed as a variant of GSP using a different screening method and
without a Rinott-like Stage~3. \NSGS\ is a parallel implementation of
the NSGS procedure \citep{Nelson2001}, and is a simplification of GSP
without the iterative screening Stage~2.

We implement all three procedures using MPI and test them on different
instances of the throughput maximization problem. We measure the
performance of these procedures in terms of total wall-clock time
and simulation replications required to find a solution, and report
them in Table~\ref{tab:GSPvNHHvNSGS}. Preliminary runs on smaller test
problems suggest that the variation in these two measures between
multiple runs of the entire selection procedure are limited. Therefore
we only present results from a single replication to save core hours.

\cite{Ni2014} argue that NHH
tends to devote excessive simulation effort to systems with means that
are very close to the best, whereas \NSGS\ has a weaker screening
mechanism but its Rinott stage can be effective when used with a large
$\delta$, which is associated with higher tolerance of an optimality
gap. GSP, by design, combines iterative screening with a Rinott
stage. Like \NSGS, we expect that GSP will cost less with a large
$\delta$ as the Rinott sample size is $O(1/\delta^2)$, but its
improved screening method should eliminate more systems than \NSGS\
before entering the Rinott stage. Therefore, we expect GSP to work
particularly well when a large number of systems exist both inside and
outside the indifference zone. This intuition is supported by the
outcomes of the medium and large test cases with $\delta=0.1$, when
GSP outperforms both NHH and \NSGS.

% We
%observe that GSP outperforms NHH and \NSGS\ in many problem instances,
%costing less wall-clock time as well as fewer simulation replications,
%despite providing a strong PGS guarantee.

%\cite{Ni2014} argue that NHH eliminates systems outside the indifference zone efficiently via iterative screening but tends to overspend simulation replications on systems within the indifference zone, 

\subsubsection{A Comparison of MPI and Hadoop Versions of GSP}

\begin{table}
    \centering
    \caption{A comparison of two implementations of GSP using
        parameters $\delta=0.1$, $n_0=50$, $\alpha_1=\alpha_2=2.5\%$,
        $\bar{r} = 1000/\beta$. ``Total time'' is summed over
        all cores. (Results to 2 significant figures)}
    \renewcommand{\arraystretch}{1}
    \begin{tabular}{lllrrrrr}
        \toprule
        Configuration & $\beta$ & Version & Number of & Wall-clock & \multicolumn{2}{c}{Total time} & Utilization\\
        %        \cmidrule(lr){6-7} 
        & & & replications & time & Simulation  & Screening & \\
        &&& ($\times 10^6$)&(sec)&($\times 10^3$ sec)&
        (sec)& \%
        \\
        \hline
        3,249 systems  & 100   & HADOOP & 0.46 & 460 & 0.34 & 0.14  & 1.2 \\
        on 64 cores &       & MPI & 0.50 & 3.0  & 0.18 & 0.01  & 94 \\
        & 200   & HADOOP & 0.63 & 280 & 0.41 & 0.10  & 2.3 \\
        &       & MPI & $0.69$ & 4.1  & 0.25 & 0.01  & 95 \\
        \hline
        57,624 systems & 100   & HADOOP & 8.8 & 550 & 5.1 & 1.9  & 15 \\
        on 64 cores &       & MPI & $9.1$ & 53 & 3.3 & 0.89  & 98 \\
        & 200   & HADOOP & 12 & 410 & 7.0 & 1.7  & 27 \\
        &       & MPI & $13$ & 75 & $4.7$ & 0.83  & 98 \\
        \hline
        1,016,127  & 100 & HADOOP & 280 & 1300 & 160 & 120 & 12 \\
        systems & & MPI & 320 & 120 & 110 & 30 & 91 \\
        on 1,024 cores& 200 & HADOOP & 340 & 810 & 190 & 89 & 23 \\
        & & MPI & 380 & 140 & 140 & 29 & 97 \\
        \bottomrule& \\ [-5ex]& %[-10ex]& 
    \end{tabular}%
    \renewcommand{\arraystretch}{1}
    \label{tab:MPIvHADOOP}%	
\end{table}%

%We observe that on Stampede, each simulation replication of the throughput maximization test problem takes $3.4\times 10^{-4}$ seconds on average when the code is compiled using the Intel C++ compiler with MVAPICH2. On the other hand, the same code re-written in Java runs each replication in an average of $5.8\times 10^{-4}$ seconds running JDK 1.7.0 and Apache Hadoop 1.2.1. While the communication time between master and the worker is estimated to take $10^{-6}$ to $10^{-4}$ each time in Section~\ref{sect:implementation_mpi}, communication in MapReduce are handled automatically and harder to measure directly. 

%As both methods implement the GSP, we should expect them to cost roughly the same number of simulation replications for each problem instance. 

We now focus on GSP and test its two implementations discussed in \S\ref{sect:implementation_parallel}. Since Stage~0 is not included in the MapReduce implementation, we also remove it from the MPI version to have a fair comparison.
Both procedures are tested on Stampede. While the cluster features highly optimized C++ compilers and MPI implementations, it provides relatively less support for MapReduce. Our MapReduce jobs are deployed using the myhadoop software \citep{MYHADOOP}, which sets up an experimental Hadoop environment on Stampede.

{
    Another difference is that we perform less screening in MPI than in Hadoop. In our initial experiments, we observed that the master could become overwhelmed by communication with the workers in the screening stages, and we fixed this problem by screening using only the $20$ best systems from other cores, versus the best systems from {\em all} other cores in Hadoop. While less screening is not a non-negligible effect, it will be apparent in our results that it is dominated by the time spent with simulation.}

Before we proceed to the results, we define core utilization, an important measure of interest, as 
\begin{align*}
\text{Utilization}=\frac{\text{total time spent on simulation}}{\text{wall-clock time }\times\text{ number of cores}}.
\end{align*}
Utilization measures how efficiently the implementations use the available cores to generate simulation replications. The higher the utilization, the less overhead the procedure spends on communication and screening.

In Table~\ref{tab:MPIvHADOOP} we report the number of simulation
replications, wall-clock time, and utilization for each of the GSP
implementations. The MPI implementation takes substantially less
wall-clock time than MapReduce to solve every problem instance,
although it requires slightly more replications due to its asynchronous
and distributed screening. 
The gap in wall clock times narrows as the batch size $\beta$ and/or the system-to-core ratio are increased.
Similarly, the MPI implementation also yields much higher utilization, spending more than 90\% of the total computation time on simulation runs in all problem instances. 
%In the largest instance, the drop in utilization of the MPI implementations is due to the increase in the number of cores employed, which results in a more congested master (if we inflate the burn-in period of the simulation run by 10 times, effectively reducing the frequency of communication by around 90\%, then the overall utilization for the largest problem instance rises to above 95\%). 
Compared to the MPI implementation, the MapReduce version utilizes core hours less efficiently but again its utilization significantly improves as we double batch size and increase the system-to-core ratio.

\begin{figure}
    \centering
    {
        \includegraphics*[width=.7\textwidth]{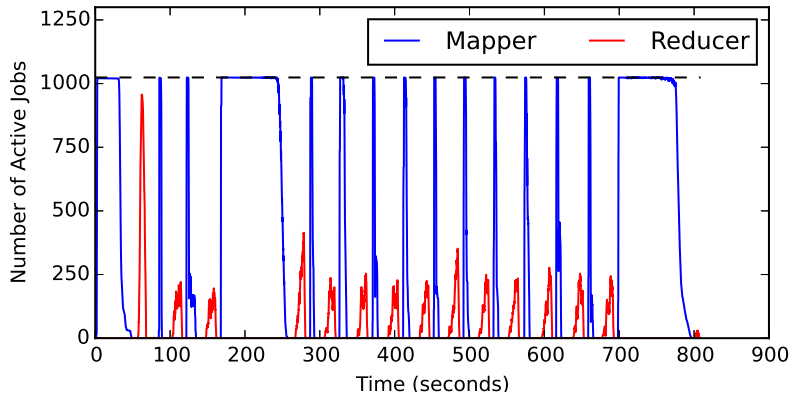}
    }
    \caption{A profile of a MapReduce run solving the largest problem instance with $k=1,016,127$ on 1024 cores, using parameters $\alpha_1=\alpha_2=2.5\%$, $\delta=0.1$, $\beta=200$, $\bar{r}=5$. \label{fig:hadoop_profile}}
    %{More explanations here.}
    {}
\end{figure}

To further understand the low utilization, we give the number of
active Mapper and Reducer jobs over an entire MapReduce run in
Figure~\ref{fig:hadoop_profile}. The plot reveals a number of reasons
for low utilization. First, there are non-negligible gaps between Map
and Reduce phases, which are due to an intermediary ``Shuffle'' step
that collects and sorts the output of the Mappers and allocates it to
the Reducers. Second, as the amount of data shuffled is likely to
vary, the Reducers start and finish at different times. Third, owing
to the varying amount of computing required for different systems,
some Mappers take longer than others. In all, the strictly
synchronized design of Hadoop causes some amount of core idleness that
is perhaps inherent in the methodology, and therefore unavoidable.
Nevertheless, the fact that utilization increases as average batch size $\beta$ or the system-to-core ratio increases suggests that the Hadoop overhead becomes less pronounced as the amount of computation work per Mapper increases. Therefore we expect utilization to also improve and become increasingly competitive with MPI's for problems that feature a larger solution space or longer simulation runs.

\subsubsection{Robustness to Unequal and Random Run Times}

%Intuitively, the fact that our simulation times are not highly variable relative to their means is why MapReduce is competitive with MPI, since this presumably reduces the impact of synchronization delays. On the other hand, for settings where the simulation times exhibit enough variation that one simulation takes much longer than the others, one would expect that the synchronization-free design of our MPI procedure would become a decisive advantage.

The MapReduce implementation allocates approximately equal numbers of
simulation replications to each Mapper and the simulation run times
per replication are nearly constant for our test problem, so the computational workload in each MapReduce iteration should be fairly balanced. Indeed, in Figure~\ref{fig:hadoop_profile} we observe that Mapper jobs terminate nearly simultaneously, which suggests that load-balancing works well.
However, if the simulation run times exhibit enough variation that one
Mapper takes much longer than the others, then we would expect
synchronization delays that would greatly reduce utilization.

\begin{table}
    \centering
    \caption{A comparison of GSP implementations using a random number
        of warm-up job releases distributed like $\min\{\exp(X),
        20,000\}$ , where $X \sim N (\mu,\sigma^2)$. We use parameters
        $\delta=0.1$, $n_0=50$, $\alpha_1=\alpha_2=2.5\%$, $\beta=200$,
        $\bar{r} = 5$. (Results to 2 significant figures)}
    \renewcommand{\arraystretch}{1}
    \begin{tabular}{lrrlrrr}
        \toprule
        Configuration & $\mu$ & $\sigma^2$ & Version &  Wall-clock time (sec) & Utilization \% \\
        %		& & & & time (sec) & \% \\
        \hline
        %\hline
        3,249 systems & 7.4 & 0.5 & HADOOP & 280 & 2.3 \\
        on 64 cores & & & MPI & 4.2 & 94 \\
        %\hline
        & 6.6 & 2.0 & HADOOP & 280 & 2.0 \\
        & & & MPI & 4.0 & 93 \\
        \hline
        57,624 systems & 7.4 & 0.5 & HADOOP & 400 & 27 \\
        on 64 cores & & & MPI & 74 & 98 \\
        %\hline
        & 6.6 & 2.0 & HADOOP & 400 & 26 \\
        & & & MPI & 70 & 98 \\
        \hline
        1,016,127 systems & 7.4 & 0.5 & HADOOP & 850 & 25 \\
        on 1,024 cores & & & MPI & 150 & 97 \\
        %\hline
        & 6.6 & 2.0 & HADOOP & 850 & 22 \\
        & & & MPI & 150 & 97 \\
        \bottomrule\\ [-5ex]
    \end{tabular}%\\ [-10ex]
    \renewcommand{\arraystretch}{1}
    \label{tab:MPIvHADOOP_uneven}%
\end{table}

%We attempted to verify this conjecture with a different suite of
%experiments, where we artificially injected variance into our run times
%by randomizing the number of job releases in the warm-up stage
%(previously, this was deterministically set to $2000$). We chose a
%truncated log-normal distribution from which to sample, as its heavy tails
%should capture the scenario where a single simulation's very large run
%time slows down the entire system. The truncation of the log-normal
%was necessary due to the fact that Hadoop has an in-built time-out
%value: after a mapper calculation has exceeded this large time value,
%it is declared a time-out by the Hadoop master. Parameters of the
%truncated lognormal distribution are given in
%Table~\ref{tab:MPIvHADOOP_uneven}.

To verify this conjecture, we design additional computational
experiments where variability in simulation run times is introduced by
warming up each system for a random number $W$ of job releases (by
default, we use a fixed 2,000 job releases in the warm-up stage). We
take $W$ to be (rounded) log-normal, parameterized so that the average
warm-up period is approximately 2,000, in the hope that the heavy
tails of the log-normal distribution will lead to occasional large run
times that might slow down the entire procedure. We also truncate the
log-normal distributions from above at 20,000 job releases to avoid
exceeding a built-in timeout limit in Hadoop. Parameters of the
truncated log-normal distribution and the results of the experiment are
given in Table~\ref{tab:MPIvHADOOP_uneven}.

We observe very similar wall-clock time and utilization in all
instances compared to the base cases in Table~\ref{tab:MPIvHADOOP}
where we used fixed warm-up periods. Both implementations seem quite
robust against the additional randomness in simulation times, despite
our intuition that the MapReduce version might be noticeably impacted
due to additional synchronization waste. A potential explanation is
that as each core is allocated at least 50 systems and each system is
simulated for an average of 200 replications in each step, the
variation in single-replication completion times is averaged
out. Rather extreme variations would be required for MapReduce to
suffer a sharp performance decrease. For problems with much longer
simulation times and a lower systems-to-core ratio, the averaging
effect might not completely cancel the variations across simulation
run times.

%Contrary to our intuition, we were not able to observe a stronger divergence between MPI and MapReduce than our base case, as both wall-clock times and utilizations remained largely the same. A potential explanation is that batching averages out the variations across individual replications, and indeed we have observed that increasing the variance of the sampling distribution has little effect in terms of slowing down MapReduce. This suggests that, in fact, our procedure is quite robust to uneven simulation run times and that rather extreme variations would be required for MapReduce to suffer a sharp performance decrease.

\section*{Acknowledgements}
This work was partially supported by NSF grant CMMI-1200315, and used the Extreme Science and Engineering Discovery Environment (XSEDE), which is supported by National Science Foundation grant number ACI-1053575.

\bibliographystyle{ormsv080}
\bibliography{smaster,sim,Ni_Sim}

\newpage

\doublespace
\numberwithin{table}{section}
\numberwithin{figure}{section}
\begin{appendices}
    \noindent\textbf{{\Large Appendices}}
    
    \renewcommand{\thesection}{A.\arabic{section}}
    
    \section{Proof of Theorem \ref{theorem:GoodSelection}.}
    \label{sect:GSPproof}
    Proving Theorem~\ref{theorem:GoodSelection} requires the following lemmas, where we use $B_\Delta(\cdot)$ to denote a Brownian motion with drift $\Delta$ and volatility one. 
    \begin{lemma} (\citealt[Theorem~1]{Hong2006})
        \label{lemma:HongBMdistribution}
        Let $m(r)$ and $n(r)$ be arbitrary nondecreasing integer-valued functions of $r=0,1,\ldots$ and $i$, $j$ be any two systems. Define $Z(m,n):=\left[\sigma_i^2/m + \sigma_{j}^2/n\right]^{-1} [ \Xbar_i(m) - \Xbar_{j}(n)]$ and $Z'(m,n):=B_{\mu_i-\mu_{j}}([\sigma_i^2/m + \sigma_{j}^2/n]^{-1})$. Then the random sequences $\{Z(m(r),n(r)):r=0,1,\ldots\}$ and $\{Z'(m(r),n(r)):r=0,1,\ldots\}$ have the same joint distribution.
    \end{lemma}
    
    %\begin{lemma} [\citeNP{Hall1997}, Equation (4)]
    %BM exit probability 
    %\end{lemma}
    
    \begin{lemma} 
        \label{lemma:lemma:bound_a_t}
        Let $i\ne j$ be any two systems. Define $\tilde{a}_{ij}(\bar{r}):= \min\{S^2_i/\sigma^2_i,S^2_{j}/\sigma^2_{j} \}a_{ij}(\bar{r})$ and $\tilde{t}_{ij}(\bar{r}):= \min\{S^2_i/\sigma^2_i,S^2_{j}/\sigma^2_{j} \}\tau_{ij}(\bar{r})$. It can be shown \citep{Hong2006} that $\min\{S^2_i/\sigma^2_i,S^2_{j}/\sigma^2_{j} \} \le {t_{ij}(r)}/{\tau_{ij}(r)}$ for all $r\ge 0 $ regardless of the sampling rules $n_i(\cdot)$ and $n_{j}(\cdot)$. Therefore
        %\begin{alignat}{3}
        $
        \tilde{a}_{ij}(\bar{r})
        \le {t_{ij}(r)}a_{ij}(\bar{r})/{\tau_{ij}(r)} 
        \text{ and }
        \tilde{t}_{ij}(\bar{r})
        \le {t_{ij}(r)}\tau_{ij}(\bar{r})/{\tau_{ij}(r)}
        \label{eq:tilde_a}
        $
        %\\
        %\tilde{t}
        %%&:= \min\{S^2_i/\sigma^2_i,S^2_{j}/\sigma^2_{j} \} 
        %%&
        %\le \frac{t_{ij}(r)}{\tau_{ij}(r)}t \label{eq:tilde_t}
        %\end{alignat}
        regardless of the sampling rules $n_i(\cdot)$ and $n_{j}(\cdot)$ for all $r\ge 0$.
    \end{lemma}
    
    %\begin{lemma} [\cite{Jennison1980}, Appendix 3] \label{lemma:JJTdiscreteprocess}
    %For any symmetric continuation region $C$ and any $\Delta\ge 0$, consider two processes: a Brownian motion $B_\Delta(\cdot)$, and a discrete process obtained by observing $B_\Delta(\cdot)$ at a random, increasing sequence of times $\{t_i:i=0,1,2,\ldots\}$ where $t_0=0$ and the value of $t_i$ for $i>0$ depends on $B_\Delta(\cdot)$ only through its value in the period $[0,t_{i-1}]$. Define $\tau^C=\inf\{ t'>0: B_\Delta(t')\not\in C \}$ and $\tau^D=\inf\{ t_i: B_\Delta(t_i)\not\in C \}$. Then $\Pr [B_\Delta(\tau^D) < 0 ] \le \Pr [B_\Delta(\tau^C) < 0 ]$.
    %\end{lemma}
    
    \begin{lemma} (\citealt[Lemma~4]{Hong2006}) \label{lemma:smallContRegion}
        Let $g_1(\cdot)$, $g_2(\cdot)$ be two non-negative-valued functions such that $g_2(t')\ge g_1(t')$ for all $t'\ge 0$. Define symmetric continuations $C_m:=\{(t',x):-g_m(t')\le x \le g_m(t') \}$ and let $T_m:=\inf\{t':B_\Delta(t')\not\in C_m\}$ for $m=1,2$. If $\Delta\ge 0$, then $P[B_\Delta(T_1)<0]\ge P[B_\Delta(T_2)<0]$.
    \end{lemma}
    
    \begin{lemma} 
        %(\cite{Hall1997})
        \label{lemma:HallErrorProb}
        %Let $C=\{ (t',x): t'\le t \text{ and } -a\le x\le a \}$ and $T_0=\inf \{t':B_0(t')\not\in C \}$. Then $P[B_0(T_0) <0\text{ and } T_0< t] = \sum_{s=0}^{\infty} 2(-1)^s \bar{\Phi}({a(2s+1)}/{\sqrt{t}}) $. 
        By the reflection principle of Brownian motion,
        $P[\min_{0\le t'\le t}B_0(t')<-a]=2P[B_0(t)<-a]=2\bar{\Phi}(a/\sqrt{t})$ for all $a,t>0$.
    \end{lemma}
    
    \begin{lemma} (\citealt{Tamhane1977}) \label{lemma:TamhaneIneq}
        Let $V_1, V_2,\ldots, V_k$ be independent random variables, and let $G^w (v_1 , v_2 ,\ldots , v_k )$, $j = 1, 2, \ldots, p$, be non-negative, real-valued functions, each one nondecreasing in each of its arguments. Then
        \[
        E\left[\prod_{j=1}^p G^w(V_1 , V_2 , \ldots, V_k )\right]\ge \prod_{j=1}^{p} E[G^w(V_1, V_2 , \ldots, V_k)].
        \]
    \end{lemma}
    
    \begin{lemma} (After \citealt{NM1995} and \citealt[Lemma~1]{Nelson2001}) \label{lemma:secondstagePGS}
        For any $G_2\subseteq \mathcal S$, Stage~3 guarantees to select a system $K\in G_2$ such that $\Pr \left[\max_{i\in G_2} \mu_i-\mu_K\le\delta\right]\ge 1-\alpha_2$. 
        If, in addition, Stages~1~and~2 jointly guarantee that $\Pr[k\in G_2]\ge 1-\alpha_1$, then $$\Pr \left[\text{The procedure selects system }K: \mu_k-\mu_K\le \delta\right] \ge 1-\alpha_1-\alpha_2.$$
    \end{lemma}

    \proof
    {Proof of Theorem~\ref{theorem:GoodSelection}}
    
    %
    %For any $i\le k-1$, let 
    %\begin{align*}
    %R_1&=\text{The first } r \text{ such that }|Y_{ki}(\tau_{ki}(r))| >a,
    %\\
    %R_2&=\text{The first } r \text{ such that }|B_{\mu_k-\mu_i}(t_{ki}(r))| >
    %\frac{t_{ki}(r)}{\tau_{ki}(r)}a,
    %\\
    %R_3&=\text{The first } r \text{ such that }|B_{\mu_k-\mu_i}(t_{ki}(r))| >
    %\tilde{a}, \text{ and}
    %\\
    %T_4&=\text{The first } t \text{ such that }|B_{\mu_k-\mu_i}(t)| >
    %\tilde{a}.
    %\end{align*}
    
    First, note that for any system~$i$, it is well known \citep[page 218]{Casella2002} that $\bar{X}_i(n_1)|S_i^2$ is normally distributed and $X_{i\ell}$ is independent of $S_i^2$ for all $\ell>n_1$. Furthermore, $\bar{T}_i$ is obtained in Stage~0 independently of all $X_{i\ell}$'s. Therefore, choosing the sampling rule based on $\bar{T}_i$ and $S_i^2$ does not affect the normality of the $\{\bar{X}_i(n_i(r)):r=0,1,\ldots,\bar{r}\}$ sequence. 
    
    For any two systems $i$ and $j$, let $KO_{ij}$ be the event that system $i$ eliminates system $j$ in Stages~1~or~2. 
    It then follows that 
    \begin{align*}
    \Pr&[KO_{ik} \text{ in Stages~1~or~2}]
    &\\=&E[\Pr[KO_{ik} \text{ in Stages~1~or~2}|S^2_k,S^2_i]]
    &\\\le&E[\Pr[Y_{ki}(\tau_{ki}(r))<-a_{ij}(\bar{r}) 
    %\text{ and } \tau_{ki}(r)< t 
    \text{ for some } r \le \bar{r}|S^2_k,S^2_i]] 
    \\&\text{ since system $i$ could be eliminated by some other system before it can eliminate system $k$}
    &\\=&E[\Pr[Y_{ki}(\tau_{ki}(r))<-a_{ij}(\bar{r}) 
    \text{ and } \tau_{ki}(r)\le \tau_{ki}(\bar{r})
    \text{ for some } r |S^2_k,S^2_i]]
    &\\=&E[\Pr[Z_{ki}(t_{ki}(r)) <-\frac{t_{ki}(r)}{\tau_{ki}(r)}a_{ij}(\bar{r}) 
    \text{ and } t_{ki}(r)\le \frac{t_{ki}(r)}{\tau_{ki}(r)}\tau_{ki}(\bar{r}) \text{ for some } r|S^2_k,S^2_i]]
    &\\=&E[\Pr[B_{\mu_k-\mu_i}(t_{ki}(r)) < -\frac{t_{ki}(r)}{\tau_{ki}(r)}a_{ij}(\bar{r}) \text{ and } t_{ki}(r)\le \frac{t_{ki}(r)}{\tau_{ki}(r)}\tau_{ki}(\bar{r}) \text{ for some } r|S^2_k,S^2_i]]
    \text{ by Lemma~\ref{lemma:HongBMdistribution}} 
    \\\le&E[\Pr[B_{\mu_k-\mu_i}(t_{ki}(r)) < -\tilde{a}_{ij}(\bar{r}) \text{ and } t_{ki}(r)\le\tilde{t}_{ij}(\bar{r}) \text{ for some } r|S^2_k,S^2_i]]
    \text{ by Lemmas~\ref{lemma:lemma:bound_a_t}~and~\ref{lemma:smallContRegion}
        %~and~\ref{lemma:JJTdiscreteprocess}
    }
    %\\\le&E[\Pr[B_{\mu_k-\mu_i}(T_4) < -\tilde{a} \text{ and } T_4<\tilde{t}|S^2_k,S^2_i]]
    %\text{ by Lemma~\ref{lemma:smallContRegion} } 
    %\\\le&E[\Pr[B_{\mu_k-\mu_i}(t') < -\tilde{a} \text{ for some } t': t'<\tilde{t}|S^2_k,S^2_i]]
    %\text{ by Lemmas~\ref{lemma:lemma:bound_a_t} and \ref{lemma:smallContRegion} } 
    \\\le&E[\Pr[B_{\mu_k-\mu_i}(t) < -\tilde{a}_{ij}(\bar{r}) \text{ for some } t\le\tilde{t}_{ij}(\bar{r}) |S^2_k,S^2_i]]
    \\\le&E[\Pr[B_{0}(t) < -\tilde{a}_{ij}(\bar{r}) \text{ for some } t\le\tilde{t}_{ij}(\bar{r})|S^2_k,S^2_i]]
    \text{ since } \mu_k\ge \mu_i
    \\=& 
    %\eric{
    E\left[2\bar{\Phi}\left(\frac{\tilde{a}_{ij}(\bar{r})}{\sqrt{\tilde{t}_{ij}(\bar{r})}}\right)\right]
    %}
    \text{ by Lemma~\ref{lemma:HallErrorProb}}
    \\=& E\left[2\bar{\Phi}\left(\frac{a_{ij}(\bar{r})}{\sqrt{\tau_{ij}(\bar{r})(n_1-1)}}\sqrt{\min \left\lbrace \frac{(n_1-1)S_i^2}{\sigma^2_i},\frac{(n_1-1)S_k^2}{\sigma^2_k} \right\rbrace}\right)\right]
    \\=& E\left[2\bar{\Phi}\left(\eta\sqrt{\min \left\lbrace \frac{(n_1-1)S_i^2}{\sigma^2_i},\frac{(n_1-1)S_k^2}{\sigma^2_k} \right\rbrace}\right)\right] \text{ by choice of $a_{ij}(\bar{r})$}
    \\=&1-(1-\alpha_1)^{\frac{1}{k-1}}
    \mbox{ by \eqref{eq:Choice_eta}, since } {(n_1-1)S_i^2}/{\sigma^2_i} \text{ and } {(n_1-1)S_k^2}/{\sigma^2_k} \text{ are i.i.d. } \chi^2_{n_1-1} \text{ random variables}.
    \end{align*}
    Then, noting that simulation results from different systems are mutually independent, we have
    \begin{align*}
    \Pr [\text{system } k \in {\mathcal S}_1] 
    &= E\left[\Pr \left\lbrace\bigcap_{i=1}^{k-1} \overline{KO}_{ik}|X_{k1},X_{k2},\ldots\right\rbrace\right]
    \\&=E\left[ \prod_{i=1}^{k-1} \Pr \left\lbrace \overline{KO}_{ik}|X_{k1},X_{k2},\ldots\right\rbrace \right]
    \\&\ge \prod_{i=1}^{k-1} E\left[ \Pr \left\lbrace \overline{KO}_{ik}|X_{k1},X_{k2},\ldots\right\rbrace \right]
    \text{ by Lemma~\ref{lemma:TamhaneIneq} }
    \\&= \prod_{i=1}^{k-1} \Pr\left[ \overline{KO}_{ik} \right]
    \ge \prod_{i=1}^{k-1} \left[1-\left(1-(1-\alpha_1)^{\frac{1}{k-1}}\right)\right] = 1-\alpha_1.
    \end{align*}
    
    Finally, we invoke Lemma~\ref{lemma:secondstagePGS} to complete the proof. 
    \endproof

    \section{Full Description of the MPI implementation} \label{sect:FullImplementation_MPI}

    \SetAlFnt{\scriptsize}
    \begin{figure}[]
        %\adjustbox{valign=t}{
        \begin{minipage}[t]{.5\linewidth}
            \begin{algorithm}[H]
                {{\underline{\textbf{Master Core Routine}}}}
                
                \KwIn{List of systems $\mcS$; %Configurations to simulate each $i\in\mcS$;
                    %Stage~0 size $n_0$; 
                    Average Stage~2 batch size $\beta$; Parameters $\delta,\alpha_1,\alpha_2,n_0,n_1,\bar{r}$ and a random number \textit{seed}.}
            \end{algorithm}
        \end{minipage}
        %}%
        %\adjustbox{valign=t}{
        \begin{minipage}[t]{.5\linewidth}
            \begin{algorithm}[H]{{\underline{\textbf{Worker Core Routine}}}}
                
                \KwIn{List of systems $\mcS$; 
                    %Stage~0 size $n_0$; 
                    Parameters $\delta,\alpha_1,\alpha_2,n_0,n_1$.
                    $\phantom{\alpha_0 \beta_0 \alpha \beta \alpha \beta \alpha \beta \alpha \beta}$}
            \end{algorithm}
        \end{minipage}
        %}%
        % % % % % % % % % % % % % % % % % % % % % % % % % % %
        % % % % % % % % % % % % % % % % % % % % % % % % % % %
        % Prep
        % % % % % % % % % % % % % % % % % % % % % % % % % % %
        % % % % % % % % % % % % % % % % % % % % % % % % % % %
        
        \adjustbox{valign=t}{
            \begin{minipage}[t]{.5\linewidth}
                \begin{algorithm}[H]
                    \Begin(\textbf{Preparation}: Setting up random number streams){ 
                        Initialize random number generator using the \textit{seed}\;
                        \ForEach{worker $w=1,2,\ldots,c$}{ 
                            Generate a new random number stream $U_w$\;
                            Send $U_w$ to $w$\;
                        }
                    }
                \end{algorithm}
            \end{minipage}}%
            \adjustbox{valign=t}{
                \begin{minipage}[t]{.5\linewidth}
                    \begin{algorithm}[H]
                        \Begin(\textbf{Preparation}: Setting up random number streams){ 
                            Receive random number stream $U_w$\;
                            Initialize random number generator using $U_w$\;
                        }
                    \end{algorithm}
                \end{minipage}}%
                
                % % % % % % % % % % % % % % % % % % % % % % % % % % %
                % % % % % % % % % % % % % % % % % % % % % % % % % % %
                % stage 0
                % % % % % % % % % % % % % % % % % % % % % % % % % % %
                % % % % % % % % % % % % % % % % % % % % % % % % % % %
                
                \adjustbox{valign=t}{
                    \begin{minipage}[t]{.5\linewidth}
                        \begin{algorithm}[H]
                            \Begin(\textbf{Stage 0}: Estimating simulation completion time){ 
                                %Randomly permute systems in $\mcS$\;
                                %Partition $\mcS$ into $\{G_w:w=1,2,\ldots,c\}$\;
                                $\{G_w^0:w=1,2,\ldots,c\}\leftarrow$\texttt{Partition}$(\mcS,0)$\;
                                \ForEach{worker $w=1,2,\ldots,c$}{
                                    Send $G_w^0$ to Worker~$w$\;
                                    %\For{$i=1$ to $|\mcS|/|\mcW|$}{
                                    %Send$(j, n_0)$\;
                                    %} 
                                }
                                \texttt{Collect}($\overline{T}_i$)\;
                            }
                        \end{algorithm}
                    \end{minipage}}%
                    \adjustbox{valign=t}{
                        \begin{minipage}[t]{.5\linewidth}
                            \begin{algorithm}[H]
                                \Begin(\textbf{Stage 0}: Estimating simulation completion time){ 
                                    Receive the set of systems to simulate, $G_w^0$\;
                                    \ForEach{system $i\in G_w^0$}{
                                        \texttt{Simulate}$(i,n_0,  \text{simulation time } \overline{T}_i)$\;
                                    }
                                    Return $\{\overline{T}_i:i\in G_w^0 \}$ to master\;
                                }
                            \end{algorithm}
                        \end{minipage}}%
                        
                        % % % % % % % % % % % % % % % % % % % % % % % % % % %
                        % % % % % % % % % % % % % % % % % % % % % % % % % % %
                        %stage 1
                        % % % % % % % % % % % % % % % % % % % % % % % % % % % 
                        % % % % % % % % % % % % % % % % % % % % % % % % % % %
                        
                        \adjustbox{valign=t}{
                            \begin{minipage}[t]{.5\linewidth}
                                \begin{algorithm}[H]
                                    \Begin(\textbf{Stage 1}: Estimating sample variances){
                                        %Partition $\mcS$ into $\{G_w:w\in \mcW\}$\;
                                        $\{G_w^1:w=1,2,\ldots,c\}\leftarrow$\texttt{Partition}$(\mcS,1)$\;
                                        % such that the sum of the completion times for systems in $G_w$ is approx. $T_{\text{tot}}/|\mcW|$, where $T_{\text{tot}}$ is the total completion time across all systems estimated in Stage~0\;
                                        \ForEach{worker $w=1,2,\ldots,c$}{
                                            Send $G_w^1$ to Worker~$w$\;
                                            % Send$(j,n_1)$\;
                                        }
                                        \texttt{Collect}($S_i^2$ and $\texttt{Stat}_{i,0} $)\;
                                        %\uIf{Screen on the master level}{
                                        %$\mcS\leftarrow$\texttt{Screen}$(\mcS,0,0,false)$\;
                                        %}
                                        %\Else {
                                        $\{\mcS, G_w^1\}\leftarrow$\texttt{RecvScreen}$(w)$\;
                                        %}
                                    }
                                \end{algorithm}
                            \end{minipage}}%
                            \adjustbox{valign=t}{
                                \begin{minipage}[t]{.5\linewidth}
                                    \begin{algorithm}[H]
                                        \Begin(\textbf{Stage 1}: Estimating sample variances){ 
                                            Receive the set of systems to simulate, $G_w^1$\;
                                            \ForEach{system $i\in G_w^1$}{
                                                \texttt{Simulate}$(i,n_1, (S_i^2, \texttt{Stat}_{i,0}))$\;
                                            }
                                            Return $\{(S_i^2,  \texttt{Stat}_{i,0} ):i\in G_w \}$ to master\;
                                            %\If{Screen on the worker level}{
                                            $G_w\leftarrow$\texttt{Screen}$(G_w^1,0,0,false)$\;
                                            \texttt{SendScreen}$(G_w^1)$\;
                                            %}
                                        }
                                    \end{algorithm}
                                \end{minipage}}%
                                
                                \caption{Stages~0 and 1, MPI Implementation: Master (left) and workers (right) routines} \label{fig:master_worker_algo_1}
                            \end{figure}
                            \begin{figure}
                                % % % % % % % % % % % % % % % % % % % % % % % % % % %
                                % % % % % % % % % % % % % % % % % % % % % % % % % % %
                                %stage 2
                                % % % % % % % % % % % % % % % % % % % % % % % % % % %
                                % % % % % % % % % % % % % % % % % % % % % % % % % % %
                                
                                \adjustbox{valign=t}{
                                    \begin{minipage}[t]{.5\linewidth}
                                        \begin{algorithm}[H]
                                            \Begin(\textbf{Stage 2}: Iterative screening){
                                                $G_1\leftarrow$ systems that survived Stage~1\;
                                                $\{G_2^w:w=1,2,\ldots,c\}\leftarrow$\texttt{Partition}$(G_1,2)$\;
                                                $\mcS\leftarrow G_1$\;
                                                \ForEach{worker $w=1,2,\ldots,c$}{
                                                    Send $G_1$, $G_2^w$ to Worker~$w$\;
                                                    \ForEach{system $i\in G_1$ }{
                                                        Send $S_i^2$ %, the sample std. dev. of System~$i$ 
                                                        from Stage~1 to Worker~$w$\;
                                                    }
                                                    \ForEach{system $i\in G_2^w$ }{
                                                        %$\texttt{Stat}_{i,0}\leftarrow (\overline{Y}_{i,0}, n_1)$\;
                                                        Send $\texttt{Stat}_{i,0}$ to worker $w$\;
                                                    }
                                                }
                                                %$\left\{b_i: i\inG_1\right\}\leftarrow$\texttt{ChooseBatchSize}$\left(\overline{b}, S^2_i, \overline{T}_i \text{ for all } i\inG_1\right)$;
                                                %$r_{\text{Global}}\leftarrow 1$;  
                                                $b_i\leftarrow\texttt{BatchSize}(i,\beta) $,
                                                $q_i\leftarrow 1$ for all $i\in G_1$\; 
                                                $r_w^\text{sent}\leftarrow 0$, $r_w^\text{received}\leftarrow 0$, $r_w^\text{screened}\leftarrow 0$, $\texttt{flag}_w\leftarrow 0$ for all $w=1,2,\ldots,c$\;
                                                \While{$|\mcS|> 1$ and $r_w^\text{screened}<\bar{r}$ for some $w$}{
                                                    \nosemic Wait for the next worker~$w$ to call\; \pushline\dosemic\texttt{Communicate()}\;
                                                    \popline 
                                                    \uIf{$\texttt{flag}_w=1$} {
                                                        \tcc{Send screening task to worker $w$}
                                                        $\{i,q_i,\texttt{Stat}_{i,q_i}\}\leftarrow$\texttt{RecvOutput}$(w)$;
                                                    }
                                                    \ElseIf{\texttt{flag}$_w=2$}{
                                                        \tcc{Send simulation task to worker $w$}
                                                        $\{\mcS, G_2^w,r_w^\text{screened}\}\leftarrow$\texttt{RecvScreen}$(w)$\;
                                                        $\{i^*_w,r_w^\text{received}, \{\texttt{Stat}_{i^*_w,r}: r\le r_w^\text{received} \} \}\leftarrow$\texttt{RecvBest}$(w)$\;
                                                    }
                                                    \If {$|\mcS|>1$} {
                                                        $r^\text{current}\leftarrow $\texttt{CountBatch}$(w)$\;
                                                        \uIf{$r^\text{current}>r_w^\text{sent} $} {
                                                            $\texttt{flag}_w\leftarrow 2$; \texttt{SendAction}($w, $\texttt{flag}$_w$)\;
                                                            \texttt{SendStats}($w, r_w^\text{sent}, r^\text{current}$); \texttt{SendBestStats}$(w)$\;
                                                            $r_w^\text{sent}\leftarrow r^\text{current}$\;
                                                            %Send $\texttt{Stat}_{il}$ for all $i\in G_2^w$, $l'\ge l$ for which $\texttt{Stat}_{il'}$ has been received for all $i\in G_2^w$, to worker $w$\;
                                                            %Send stats. of the best systems to worker $w$\;
                                                            %$\texttt{flag}_w\leftarrow 2$\;
                                                        }
                                                        \Else{
                                                            $\texttt{flag}_w\leftarrow 1$; \texttt{SendAction}($w,\texttt{flag}_w$)\;
                                                            Select next $i\in \mcS$ such that $ q_{i} = q_{\text{Global}}$\;
                                                            \texttt{SendSim}$(w,i, q_{i}, b_{i})$\;
                                                            $q_{i}\leftarrow q_{i}+1$\;
                                                            \If {$q_{i}>q_{\text{Global}}$ for all $i\in \mcS$} {
                                                                $q_{\text{Global}}\leftarrow q_{\text{Global}}+1$\;
                                                            }
                                                        }
                                                    }
                                                    %Receive indexes of sys. eliminated on Worker~$w$;
                                                    %Update surviving sys.; 
                                                    %Receive stats. (sample mean and sample size) of the best system for Worker~$w$, for each batch simulated up to this point;
                                                    %Send stats. of the best sys. from other workers to Worker~$w$; 
                                                    %Instruct Worker~$w$ to continue simulation\;
                                                }
                                                Send a termination instruction to all workers\;
                                            }
                                        \end{algorithm}
                                    \end{minipage}
                                }%
                                \adjustbox{valign=t}{
                                    \begin{minipage}[t]{.5\linewidth}
                                        \begin{algorithm}[H]
                                            
                                            \Begin(\textbf{Stage 2}: Iterative screening){
                                                Receive the set of systems that survived, $ G_1$\;
                                                Receive the set of systems to screen, $G_2^w$\;
                                                \ForEach{System~$i\in  G_1$}{
                                                    Receive $S_i^2$ collected in Stage~1\;
                                                }
                                                \ForEach{system $i\in G_2^w$ }{
                                                    Receive $\texttt{Stat}_{i,0}$ from the master\;
                                                }
                                                $r_w\leftarrow 0$\;
                                                \texttt{Communicate}$()$\;
                                                \While{No termination instruction received}{
                                                    $\texttt{flag}_w\leftarrow$\texttt{RecvAction}$()$\;
                                                    \uIf {$\texttt{flag}_w=2$ } {
                                                        \nosemic $\left\lbrace r^\text{new}, \lbrace \texttt{Stat}_{i,r}: i\in G_2^w, r_w+1\le r\le r^\text{new} \rbrace  \right\rbrace$\; \pushline\dosemic $\leftarrow$\texttt{RecvStats}$()$\;
                                                        \popline  
                                                        \nosemic $\lbrace \mcW, 
                                                        \{r_{w'}^{received}: w'\in \mcW\}, $\;
                                                        \pushline 
                                                        $\{\texttt{Stat}_{i^*_{w'},r}:w'\in \mcW, r\le r_{w'}^{received} \} \rbrace$ \;
                                                        \dosemic $\leftarrow$\texttt{RecvBestStats}$()$\;
                                                        \popline$G_2^w\leftarrow$\texttt{Screen}$(G_2^w,r_w+1,r^\text{new},true)$\;
                                                        $r_w\leftarrow r^\text{new}$\;
                                                        \texttt{Communicate}$()$\;
                                                        \texttt{SendScreen}$(G_2^w,r_w)$;
                                                        \texttt{SendBest}$(r)$\;
                                                    }
                                                    \Else {
                                                        $\{i,q_i,b_i \}\leftarrow$\texttt{RecvSim}$()$\;
                                                        \texttt{Simulate}$(i, b_i, \texttt{Stat}_{i,q_i})$\;
                                                        \texttt{Communicate}$()$\;
                                                        \texttt{SendOutput}$(i,q_i, \texttt{Stat}_{i,q_i})$
                                                    }
                                                    \For{$i=1$ to $r$}{
                                                        \texttt{Simulate}$(i,n_i, \bar{X}_i )$ for one batch\;
                                                    }
                                                    Screen among $r$ sys. using current batch stats.\;
                                                    \For{each batch $k$ up to the current one}{
                                                        If batch $k$ stats. available, screen the $r$ systems simulated, against the best systems from the other workers, at batch $k$\;
                                                    }
                                                    \uIf{Master sends a terminate instruction}{
                                                        continue$\leftarrow$ false;
                                                    }\ElseIf{Master is ready to communicate} {
                                                    Report indexes of eliminated sys. to master;
                                                    Report stats. of the best system for each batch simulated up to this point;
                                                    Receive stats. of the best sys. from other workers\; 
                                                }
                                            }
                                        }
                                    \end{algorithm}
                                \end{minipage}
                            }%
                            % % % % % % % % % % % % % % % % % % % % % % % % % % %
                            
                            \caption{Stage~2, MPI Implementation: Master (left) and workers (right) routines} \label{fig:master_worker_algo_2}
                        \end{figure}
                        
                        \begin{figure}[]
                            % % % % % % % % % % % % % % % % % % % % % % % % % % %
                            % % % % % % % % % % % % % % % % % % % % % % % % % % %
                            %stage 3
                            % % % % % % % % % % % % % % % % % % % % % % % % % % % 
                            % % % % % % % % % % % % % % % % % % % % % % % % % % %
                            
                            \adjustbox{valign=t}{
                                \begin{minipage}[t]{.5\linewidth}
                                    \begin{algorithm}[H]
                                        \Begin(\textbf{Stage 3}: Rinott Stage){
                                            $G_2\leftarrow$ systems that survived Stage~2\;
                                            \uIf {$|G_2|=1$} { Report the single surviving system as the best\; 
                                            } \Else {
                                            $h\leftarrow h(1-\alpha_2,n_1, |G_2|)$\;
                                            %$\mcB\leftarrow\emptyset$\;
                                            \ForEach{system $i\in G_2$}{
                                                $N_i\leftarrow \max\{n_i(\bar{r}), \left\lceil (hS_i/\delta)^2 \right\rceil \}$\;
                                                $N_i^\text{sent}\leftarrow 0$; $N_i^\text{received}\leftarrow 0$\;
                                            }
                                            \texttt{flag}$_w\leftarrow 0$ for all $w=1,2,\ldots,c$\;
                                            \While{$N_i^\text{received}<N_i-n_i(\bar{r})$ for some $i\in G_2$} {
                                                \nosemic Wait for the next worker $w$ to call \;
                                                \pushline\dosemic \texttt{Communicate()}\;
                                                \popline\If{\texttt{flag}$_w$$=1$} {
                                                    Receive $i$, $b'_{i}$ and sample mean of the current batch\;
                                                    Merge sample mean into $\Xbar_i$\;
                                                    $N_i^\text{received}\leftarrow N_i^\text{received} + b'_i$
                                                }
                                                \If{$N_i^\text{sent}<N_i-n_i(\bar{r})$ for some $i\in G_2$} {
                                                    Find an appropriate batch size $b'_i=\min \{b_i,N_i-n_i(\bar{r})-N_i^{text
                                                        sent}  \} $ for system $i$\;
                                                    Send system $i$ and $b'_i$ to worker $w$\;
                                                    $N_i^\text{sent}\leftarrow N_i^\text{sent} + b'_i$\;
                                                    \texttt{flag}$_w$$\leftarrow 1$\;
                                                }
                                            }
                                            Report the system $i^*=\arg\max_{i\in G_2} \Xbar_i(N_i) $ as the best\;
                                        }
                                        Send a termination instruction to all workers\;
                                    }
                                \end{algorithm}
                            \end{minipage}}%
                            \adjustbox{valign=t}{
                                \begin{minipage}[t]{.5\linewidth}
                                    \begin{algorithm}[H]
                                        \Begin(\textbf{Stage 3}: Rinott Stage){ 
                                            \texttt{Communicate()}\;
                                            \While{No termination instruction received} {
                                                Receive a system $i$ and batch size $b'i$ from the master\;
                                                Simulate system $i$  for $b_i$ replications\;
                                                \texttt{Communicate()}\;
                                                Send $i$, $b'_i$ and sample mean of the $b'_i$ replications to the master\;
                                            }
                                        }
                                    \end{algorithm}
                                \end{minipage}}%
                                
                                \caption{Stage~3, MPI Implementation: Master (left) and workers (right) routines} \label{fig:master_worker_algo_3}
                            \end{figure}
                            
                            The purpose of this section is to provide additional insight into our parallel codes.
                            In Figures~\ref{fig:master_worker_algo_1} through \ref{fig:master_worker_algo_3} we demonstrate in greater
                            detail how the master core allocates and distributes systems, how
                            random number streams are created and distributed together with the
                            assigned systems to ensure independent sampling, and how simulation
                            results are communicated between cores. 
                            %For notational simplicity, we assume that the number of systems $|\mcS|$ is divisible by the number of workers $|\mcW|$, although in practice we can freely distribute the few extra systems to workers. 
                            
                            We use the following notation for some subroutines in Figures~\ref{fig:master_worker_algo_1} through \ref{fig:master_worker_algo_3}:
                            \begin{description}
                                \item \texttt{Partition}($\mcS $, \textit{Stage}) The master divides the set of systems $\mcS$ into disjoint partitions $\{G^w_{\textit{Stage}}:w=1,2,\ldots,c \}$:
                                % % % % %old
                                \\In \textit{Stage 0}, all systems are simulated for $n_0$ replications to estimate simulation completion time. The master randomly permutes $\mcS$ (in case of long runtimes for some systems that are indexed closely) and assigns approximately equal numbers of systems to each $G^w_0$. 
                                \\In \textit{Stage 1}, a fixed number $n_1$ of replications are required from each system. To balance the simulation work among workers, the master chooses $G_1^w$ such that the estimated completion time $\sum_{i\in G_1^w} n_1\bar{T}_i/n_0$ is approximately equal for all $w$.
                                % % % % %new
                                %\\In \textit{Stage 1}, all systems are simulated for $n_{0}$ replications to estimate simulation completion time and variance. The master randomly permutes $\mcS$ (in case of long runtimes for some systems that are indexed closely together) and assigns approximately equal number of systems to each $G^w$.
                                \\In \textit{Stage 2}, both simulation and screening are performed iteratively. Simulation of a system is no longer dedicated to a particular worker, and $G^w_2$ is the set of systems that worker $w$ needs to screen. To load-balance the screening work, the master assigns approximately equal numbers of systems to each $G^w_2$.
                                
                                \item \texttt{Collect}(\textit{info}) The master collects \textit{info} from all workers for all systems, in arbitrary order.
                                
                                \item \texttt{Simulate}$(i,n,\text{\textit{info}})$ Worker~$w$ simulates system~$i$ for $n$ replications and records \textit{info} using the next subsubstream in $U_w^i$.
                                
                                \item $\texttt{Stat}_{i,r}$ The batch statistics for the $r$th batch of system~$i$. This includes sample size $n_i(r)$ and sample mean $\Xbar_{i}(n_i(r))$ as described in \S\ref{sect:GoodSelectionProcedure}.
                                
                                \item \texttt{BatchSize}($i,\beta$) The master calculates batch size $b_i$ system $i$ used in Stage~2. Following the recommendation from \S\ref{sect:BatchSizes}, we let 
                                \begin{align}
                                b_i=\left \lceil \frac{S_{i} \sqrt{T_{i}}}{\frac{1}{|\mcS|}\sum_{j\in\mcS}S_{j} \sqrt{T_{j}}} \beta\right \rceil 
                                \end{align} where $\beta$ is a pre-determined average batch size.
                                
                                \item \texttt{Screen}($G^w,r_0,r_1,$ \textit{useothers}) Screen systems in $G^w$ from batches $r_0$ through $r_1$ inclusive. It can be checked that worker $w$ has received $\texttt{Stat}_{i,r}$ for all $i\in\mcS$, all $r\le r_1$ and stored the data in its memory. 
                                
                                A system~$i\in G^w$ is eliminated if there exists system~$j\in G^w_2:j\neq i$ and some $r':r_0\le r'\le r_1$ such that 
                                %\eric{$\tau_{ij}(r_w)<t$ and }
                                $r'\le \bar{r}$ and $
                                Y_{ij}(r')
                                <
                                %\min[0,
                                -a_{ij}(\bar{r})
                                %+\lambda \tau_{ij}(r_w) 
                                %]
                                $ where $Y_{ij}$ and $a_{ij}$ are defined in \S\ref{sect:GoodSelectionProcedure}. 
                                
                                %For each batch $r_0\le r\le r_1$, the core first computes \begin{align}
                                %n_i(l) = n_0+rb_i 
                                %\label{eq:SampleSize}
                                %\end{align}
                                %and 
                                %\begin{align}
                                %\overline{Y}_i(n_i(l))=\left[\sum_{l'=1}^{l}\overline{Y}_{i,l'}b_i+n_0\overline{Y}_{i,0}\right]/n_i(l) \label{eq:Ybar}
                                %\end{align} for each system $i$, 
                                %then $\tau_{ij}(l)=\left[\frac{S_i^2}{n_i(l)}+\frac{S_j^2}{n_j(l)}\right]^{-1}$ and $Y_{ij}(\tau_{ij}(l))=\tau_{ij}(l)\left[ \overline{Y}_i(n_i(l))-\overline{Y}_j(n_j(l)) \right]$ for each pair $i,j\in\mcS$ and $i\ne j$, 
                                %and finally lets \\$\mcS\leftarrow\mcS\backslash\left\lbrace i\in \mcS: Y_{ij}(\tau_{ij}(l))<\min [0,-a+\lambda\tau_{ij}(l)] \text{ for some } j\in\mcS \text{ and }j\ne i \right\rbrace$.
                                %\\
                                In addition, if \textit{useothers}$=true$ and $\mcW\ne\emptyset$, then for each $w'\in \mcW$ the worker also screens the systems in $G^w$ against system $i^*_{w'}$, the best system from worker $w'$, using batch statistics $\{\texttt{Stat}_{i^*_{w'},r'}:r'\le r_{w'} \}$ 
                                %to screen systems in $\mcS$ 
                                up to batch $\min\{ r_{w'}, r_1\}$.
                                
                                \item \texttt{SendScreen$(G^w,r_w)$} and \texttt{RecvScreen$(w)$} Worker $w$ sends $r_w$ and screening results (updated $G^w$) to the master, which then updates $G^w$ and $\mcS$ on its own memory accordingly. The master also receives $r_w$ and lets $r_w^\text{screened}\leftarrow r_w$.
                                
                                \item \texttt{Communicate}$()$ Worker sends a signal to master and waits for the master to receive the signal, before proceeding.
                                
                                \item \texttt{SendSim$(w,i,q_i,b_i)$} and \texttt{RecvSim$()$} The master instructs worker $w$ to simulate the $q_i$th batch of system~$i$, for $b_i$ replications. Worker $w$ receives $i$, $q_i$, $b_i$ from the master.
                                
                                \item \texttt{SendOutput}$(i,q_i,\texttt{Stat}_{i,q_i})$ and \texttt{RecvOutput}$(w)$ Worker $w$ sends simulation output $\texttt{Stat}_{i,q_i}$ for the $q_i$th batch of system~$i$ to the master. The master stores $\texttt{Stat}_{i,q_i}$ in memory.
                                
                                \item \texttt{SendBest$()$} and \texttt{RecvBest$(w)$} Worker $w$ sends its estimated-best system $i^*_w$ (the one in $G^w$ with the highest batch mean) to the master, together with all batch statistics for system~$i^*_w$, $\{\texttt{Stat}_{i^*_w,r}: r\le r_w \} $; the master receives $r_w$ and lets $r_w^\text{received}\leftarrow r_w$.
                                
                                \item \texttt{CountBatch}$(w)$ The master finds the largest $r^\text{current}\ge r_w$ such that $\texttt{Stat}_{i,r}$ for all $i\in G^w$, $r_w<r\le r^\text{current}$ have been received by the master.
                                
                                \item \texttt{SendAction$(w,\text{flag}_w)$} and \texttt{RecvAction$()$} The master sends an indicator $\text{flag}_w$ to worker $w$, where $\text{flag}_w=1$ indicates ``simulate a batch'' and $\text{flag}_w=2$ indicates ``perform screening''.
                                
                                \item \texttt{SendStats$(w)$} and \texttt{RecvStats$()$} The master sends $\texttt{Stat}_{i,r}$ for all $i\in G^w$, $r_w<r\le r^\text{current}$ to worker $w$; the worker receives $r^\text{current}$ and lets $r^\text{new} \leftarrow r^\text{current}$; the worker should have $\texttt{Stat}_{i,r}$ for all $i\in G^w$, $0<r\le r^\text{new}$ upon completion.
                                
                                \item \texttt{SendBestStats$(w)$} and \texttt{RecvBestStats$()$} The master computes $\mcW=\{w'\ne w: |G_{w'}|>0) \}$ and sends $\mcW$ to worker $w$; the master then sends all available batch statistics for best systems, $\{\texttt{Stat}_{i^*_{w'},r}:{w'}\in \mcW, r\le r_{w'}^\text{received} \}$, to worker $w$.
                                
                            \end{description}
                            %Send$(j,n)$ denotes the sub-process ``Send configurations for the next system to Worker~$j$; create a new random number stream and send to Worker~$j$; send run length $n$ to Worker~$j$''. Receive$(i,n_i)$ denotes the sub-process ``Receive configurations for system~$i$ from the master; receive the random number stream used for system~$i$; Receive run length for system~$i$, $n_i$''. 

                            \section{Full Description of the Hadoop implementation} \label{sect:FullImplementation_Hadoop}
                            We present in this section the full details of the MapReduce implementation of GSP. 
                            
                            Each Mapper reads a comma-separated string of varied length, denoted by $[\text{value 1}, \text{value 2},\ldots, \$\texttt{type}]$, where the last component $\$\texttt{type}$ is used to indicate the specific information captured in the string.
                            A Mapper usually runs some simulation, updates batch statistics, and generates one or more \texttt{key}$: $ \{\texttt{value}\} pairs. All pairs under the same \texttt{key} are sent to the same Reducer, which is typically responsible for screening. A Reducer may generate one or more comma-separated strings which become the input to the Mapper in the next iteration.
                            
                            Each system $i$ is coupled with \texttt{stream}$_i$ which is used by some random number generator and updated each time a random number is generated. The coupling of systems and \texttt{stream}s ensures that the random numbers generated for each system in each iteration are all mutually independent. We also assume that each system $i$ is preallocated to a particular screening group, as determined by the function \texttt{Group}$(i)$.
                            
                            The procedure begins with Steps 1-3 which implements Stage~1, then enters Stage~2 where Steps 4 and 5 are run repeatedly for a maximum of $\bar{r}$ iterations. If multiple systems survive Stage~2, the procedure runs Steps 6 and 7 to finish Stage~3.
                            
                            \begin{enumerate} [labelindent=0pt,labelwidth=\widthof{\ref{last-item}},label=\bfseries Step \arabic*.,itemindent=1em,leftmargin=!]
                                \item %step 1
                                \begin{itemize}
                                    \item    
                                    \textbf{Map}: Estimate $S_i^2$
                                    \begin{enumerate}
                                        \item[]
                                        \begin{enumerate}
                                            \item [\textit{Input}] [$i$]
                                            \item [\textit{Operation}]  Initialize \texttt{stream}$_i$ with seed $i$; Simulate system $i$ for $n_1$ replications to obtain $\Xbar_i(n_1)$ and $S_i^2$. 
                                            \item [\textit{Output}] $i$: \{$\Xbar_i(n_1)$, $S_i^2$, \texttt{stream}$_i$, \$S0\}
                                        \end{enumerate}
                                    \end{enumerate}
                                    \item
                                    \textbf{Reduce}
                                    \begin{enumerate}
                                        \item[]
                                        \begin{enumerate}
                                            \item [\textit{Input}] $i$: \{$\Xbar_i(n_1)$, $S_i^2$, \texttt{stream}$_i$, \$S0\}
                                            \item [\textit{Operation}] Calculate $\sum_i S_i$.
                                            \item [\textit{Output}] [$i$, $\Xbar_i(n_1)$, $S_i^2$, \text{stream}$_i$, \$S0]
                                        \end{enumerate}
                                    \end{enumerate}
                                \end{itemize}
                                
                                \item %step 2
                                \begin{itemize}
                                    \item \textbf{Map}: Calculate batch size
                                    \begin{enumerate}
                                        \item[]
                                        \begin{enumerate}
                                            \item [\textit{Input}] [$i$, $\Xbar_i(n_1)$ $S_i^2$, \texttt{stream}$_i$ \$S0]
                                            \item [\textit{Operation}] Calculate batch size $b_i$ using $b_i=\beta S_i/(\sum_i S_i/k)$.
                                            \item [\textit{Output}] \texttt{Group}($i$): \{$i$, $\Xbar_i(n_1)$, $n_1$, $b_i$, $S_i^2$, \texttt{stream}$_i$, \$Sim\}
                                        \end{enumerate}
                                    \end{enumerate}
                                    \item \textbf{Reduce}: Screen within a group
                                    \begin{enumerate}
                                        \item[]
                                        \begin{enumerate}
                                            \item [\textit{Input}] \texttt{Group}: \{$i$, $\Xbar_i(n_i)$, $n_i$, $b_i$, $S_i^2$, \text{stream}$_i$, \$Sim\} for all $i$ in the \texttt{Group}
                                            \item [\textit{Operation}] Screen all systems in the \texttt{Group} and find the one $i^*$ with the highest mean. 
                                            \item [\textit{Output}] [$i$, $\Xbar_i(n_i)$, $n_i$, $b_i$, $S_i^2$, \texttt{stream}$_i$, \$Sim] for each surviving system $i$, and 
                                            \newline
                                            [$i^*$, $\Xbar_{i^*}(n_{i^*})$, $n_{i^*}$, $b_{i^*}$, $S_{i^*}^2$, \$Best] for the best system $i^*$ 
                                        \end{enumerate}
                                    \end{enumerate}
                                \end{itemize}
                                
                                \item 
                                \begin{itemize}
                                    \item \textbf{Map}: Share best systems between groups
                                    \begin{enumerate}
                                        \item[]
                                        \begin{enumerate}
                                            \item [\textit{Input (1)}] [$i$, $\Xbar_i(n_i)$, $n_i$, $b_i$, $S_i^2$, \texttt{stream}$_i$, \$Sim]
                                            \item [\textit{Operation (1)}]  Simply output to \texttt{Group}($i$).
                                            \item [\textit{Output (1)}] \texttt{Group}($i$): \{$i$, $\Xbar_i(n_i)$, $n_i$, $b_i$, $S_i^2$, \texttt{stream}$_i$, \$Sim\} 
                                            
                                            \item [\textit{Input (2)}] [$i^*$, $\Xbar_{i^*}(n_{i^*})$, $n_{i^*}$, $b_{i^*}$, $S_{i^*}^2$, \$Best]
                                            \item [\textit{Operation (2)}] Output to all groups.
                                            \item [\textit{Output (2)}] \texttt{Group}: \{$i^*$, $\Xbar_{i^*}(n_{i^*})$, $n_{i^*}$, $b_{i^*}$, $S_{i^*}^2$, \$Best\} for every \texttt{Group}
                                        \end{enumerate}
                                    \end{enumerate}
                                    \item \textbf{Reduce}: {Screen against the best systems from other groups}
                                    \begin{enumerate}
                                        \item[]
                                        \begin{enumerate}
                                            \item [\textit{Input}] \texttt{Group}: \{$i$, $\Xbar_i(n_i)$, $n_i$, $b_i$, $S_i^2$, \text{stream}$_i$, \$Sim\} for all $i$ in the \texttt{Group}, and 
                                            \newline 
                                            \texttt{Group}: \{$i^*$, $\Xbar_{i^*}(n_{i^*})$, $n_{i^*}$, $b_{i^*}$, $S_{i^*}^2$, \$Best\} from every other \texttt{Group}
                                            \item [\textit{Operation}] Screen all systems in \texttt{Group} against the best systems from other groups.
                                            \item [\textit{Output}] [$i$, $\Xbar_i(n_i)$, $n_i$, $b_i$, $S_i^2$, \texttt{stream}$_i$, \$Sim] for each surviving system $i$ 
                                        \end{enumerate}
                                    \end{enumerate}
                                \end{itemize}
                                
                                \item %step 4
                                \begin{itemize}
                                    \item \textbf{Map}: Simulation
                                    \begin{enumerate}
                                        \item[]
                                        \begin{enumerate}
                                            \item [\textit{Input}] [$i$, $\Xbar_i(n_i)$, $n_i$, $b_i$, $S_i^2$, \texttt{stream}$_i$, \$Sim]
                                            \item [\textit{Operation}] Simulate system $i$ for additional $b_i$ replications, update $n_i$, $\Xbar_i(n_i)$, and \texttt{stream}$_i$.
                                            \item [\textit{Output}] \texttt{Group}($i$): \{$i$, $\Xbar_i(n_1)$, $n_1$, $b_i$, $S_i^2$, \texttt{stream}$_i$, \$Sim\}
                                        \end{enumerate}
                                    \end{enumerate}
                                    \item \textbf{Reduce}: Screen within a group.
                                    
                                    (Same as Step~2 Reduce.)
                                \end{itemize}
                                
                                \item Screen against best systems from other groups.
                                
                                (Same as Step~3.)
                                
                                \item %step 6
                                \begin{itemize}
                                    \item \textbf{Map}: Determine Rinott sample sizes
                                    \begin{enumerate}
                                        \item[]
                                        \begin{enumerate}
                                            \item [\textit{Input}] [$i$, $\Xbar_i(n_i)$, $n_i$, $b_i$, $S_i^2$, \texttt{stream}$_i$, \$Sim]
                                            \item [\textit{Operation}] Output to Reducer. 
                                            \item [\textit{Output}] $i$: \{$\Xbar_i(n_i)$, $n_i$, $S_i^2$, \texttt{stream}$_i$, \$Sim\} 
                                        \end{enumerate}
                                    \end{enumerate}
                                    \item \textbf{Reduce}
                                    \begin{enumerate}
                                        \item[]
                                        \begin{enumerate}
                                            \item [\textit{Input}] $i$: \{$\Xbar_i(n_i)$, $n_i$, $S_i^2$, \texttt{stream}$_i$, \$Sim\}
                                            \item [\textit{Operation}] Calculate Rinott sample size and divide the additional sample into batches. For each batch $j$, generate a substream \texttt{stream}$^j_i$ using \texttt{steam}$_i$.
                                            \item [\textit{Output}] [$i$, $\Xbar_i(n_i)$, $n_i$, \$S2], and 
                                            \newline
                                            for each batch $j$: [$i$, \texttt{stream}$^j_i$, (size of batch $j$), \$S3] 
                                        \end{enumerate}
                                    \end{enumerate}
                                \end{itemize}
                                
                                \item \label{last-item} %step 7
                                \begin{itemize}
                                    \item \textbf{Map}: Simulate additional batches
                                    \begin{enumerate}
                                        \item[]
                                        \begin{enumerate}
                                            \item [\textit{Input (1)}] [$i$, $\Xbar_i(n_i)$, $n_i$, \$S2]
                                            \item [\textit{Operation (1)}] Output to Reducer, since this is the batch statistics generated in Stage~2.
                                            \item [\textit{Output (1)}] $1$: \{$i$, $\Xbar_i(n_i)$, $n_i$, \$S2\} 
                                            \item [\textit{Input (2)}] [$i$, \texttt{stream}$^j_i$, (size of batch $j$), \$S3]
                                            \item [\textit{Operation (2)}] Simulate batch $j$ of system $i$ for the given batch size using \texttt{stream}$^j_i$, calculate batch sample mean $\Xbar_i^j$.
                                            \item [\textit{Output (2)}] $1$: \{$i$, $\Xbar_i^j$, (size of batch $j$), \$S3 \}
                                        \end{enumerate}
                                    \end{enumerate}
                                    \item \textbf{Reduce}: Merge batches and find the best system
                                    \begin{enumerate}
                                        \item[]
                                        \begin{enumerate}
                                            \item [\textit{Input}] (This step has only one Reducer)
                                            \newline 
                                            1: \{$i$, $\Xbar_i(n_i)$, $n_i$, \$S2\} and 
                                            \newline
                                            1: \{$i$, $\Xbar_i^j$, (size of batch $j$), \$S3\} for all system $i$ and all batch $j$ 
                                            \item [\textit{Operation}] For each system $i$, merge all batches (including the one from Stage~2) to form a single sample mean.
                                            \item [\textit{Output}] Report the system $i^*$ that has the highest sample mean. 
                                        \end{enumerate}
                                    \end{enumerate}
                                \end{itemize}
                                
                            \end{enumerate}
                            
                            %\input{fig_hadoop}
                            
                            %\begin{figure}[t]
                            %\begin{center}
                            %\includegraphics[height=1.5in]{Sample-Figure}
                            %\caption{Production Possibilities Frontier Again.} \label{ECfrontier}
                            %\end{center}
                            %\end{figure}

                            % Appendix here
                            % Options are (1) APPENDIX (with or without general title) or
                            % (2) APPENDICES (if it has more than one unrelated sections)
                            % Outcomment the appropriate case if necessary
                            %
                            % \begin{APPENDIX}{<Title of the Appendix>}
                            % \end{APPENDIX}
                            %
                            % or
                            %
                            % \begin{APPENDICES}
                            % \section{<Title of Section A>}
                            % \section{<Title of Section B>}
                            % etc
                            % \end{APPENDICES}

                            % Acknowledgments here
                            %\ACKNOWLEDGMENT{NSF, XSEDE}

                            % References here (outcomment the appropriate case)
                            
                            % CASE 1: BiBTeX used to constantly update the references
                            % (while the paper is being written).
                            %\bibliographystyle{ormsv080} % outcomment this and next line in Case 1
                            %\bibliography{smaster,sim,Ni_sim} % if more than one, comma separated
                            
                            % CASE 2: BiBTeX used to generate mypaper.bbl (to be further fine tuned)
                            %\input{mypaper.bbl} % outcomment this line in Case 2

                        \end{appendices}
\end{document}